\documentclass[sigconf,screen,nonacm]{acmart} % or whatever your option is

\usepackage{amsmath,amssymb,amsfonts}
\usepackage{graphicx}
\usepackage{subcaption} %use more supported subcaption package
\usepackage{textcomp}
\PassOptionsToPackage{table,dvipsnames}{xcolor}
\usepackage{siunitx}
\usepackage{threeparttable}
\usepackage{booktabs}
\usepackage{multirow}
\usepackage{pgfplots}
\pgfplotsset{compat=1.18}
\usepackage{balance}
\usepackage{tikz}
\usepackage{fancyhdr}
\usepackage{pifont}
\usepackage{textcomp}
\usepackage{tikz}
\usepackage{xcolor}
\newcommand*\solidcircle[1]{%
  \tikz[baseline=(char.base)]\node[draw, circle, fill=black, text=white, inner sep=0.3pt, minimum size=0.5em] (char) {\textbf{\scriptsize #1}};}

\tikzset{every picture/.style={/utils/exec={\sffamily}}}

\usepackage{pgfplots,pgfplotstable}
 \usepackage[ruled, vlined, linesnumbered, commentsnumbered, longend]{algorithm2e}
\hypersetup{
    colorlinks=true,
    linkcolor=blue,
    filecolor=magenta,      
    urlcolor=cyan,
    citecolor=brown,
    pdfpagemode=FullScreen,
    }

\usetikzlibrary{patterns} % fig 12
\usetikzlibrary{positioning, arrows.meta} %fig 11
\usetikzlibrary{matrix,fit} %fig 11,12
\usepackage{listings}% fig 12
\usetikzlibrary{decorations.pathreplacing,tikzmark} % fig 12
\usetikzmarklibrary{listings} % fig 12
\definecolor{pairedOneLightBlue}{HTML}{a6cee3}
\definecolor{pairedTwoDarkBlue}{HTML}{1f78b4}
\definecolor{pairedThreeLightGreen}{HTML}{b2df8a}
\definecolor{pairedFourDarkGreen}{HTML}{33a02c}
\definecolor{pairedFiveLightRed}{HTML}{fb9a99}
\definecolor{pairedSixDarkRed}{HTML}{e31a1c}
\definecolor{butter1}{rgb}{0.988,0.914,0.310}
\definecolor{butter2}{rgb}{0.929,0.831,0.000}
\definecolor{butter3}{rgb}{0.769,0.627,0.000}
\definecolor{orange1}{rgb}{0.988,0.686,0.243}
\definecolor{orange2}{rgb}{0.961,0.475,0.000}
\definecolor{orange3}{rgb}{0.808,0.361,0.000}
\definecolor{chocolate1}{rgb}{0.914,0.725,0.431}
\definecolor{chocolate2}{rgb}{0.757,0.490,0.067}
\definecolor{chocolate3}{rgb}{0.561,0.349,0.008}
\definecolor{chameleon1}{rgb}{0.541,0.886,0.204}
\definecolor{chameleon2}{rgb}{0.451,0.824,0.086}
\definecolor{chameleon3}{rgb}{0.306,0.604,0.024}
\definecolor{skyblue1}{rgb}{0.447,0.624,0.812}
\definecolor{skyblue2}{rgb}{0.204,0.396,0.643}
\definecolor{skyblue3}{rgb}{0.125,0.290,0.529}
\definecolor{plum1}{rgb}{0.678,0.498,0.659}
\definecolor{plum2}{rgb}{0.459,0.314,0.482}
\definecolor{plum3}{rgb}{0.361,0.208,0.400}
\definecolor{scarletred1}{rgb}{0.937,0.161,0.161}
\definecolor{scarletred2}{rgb}{0.800,0.000,0.000}
\definecolor{scarletred3}{rgb}{0.643,0.000,0.000}
\definecolor{aluminium1}{rgb}{0.933,0.933,0.925}
\definecolor{aluminium2}{rgb}{0.827,0.843,0.812}
\definecolor{aluminium3}{rgb}{0.729,0.741,0.714}
\definecolor{aluminium4}{rgb}{0.533,0.541,0.522}
\definecolor{aluminium5}{rgb}{0.333,0.341,0.325}
\definecolor{aluminium6}{rgb}{0.180,0.204,0.212}

\definecolor{blind_safe_one_scheme_three_colors}{RGB}{102,194,165}
\definecolor{blind_safe_two_scheme_three_colors}{RGB}{252,141,98}
\definecolor{blind_safe_three_scheme_three_colors}{RGB}{141,160,203}

\definecolor{blind_safe_one_scheme_four_colors}{RGB}{166,206,227}
\definecolor{blind_safe_two_scheme_four_colors}{RGB}{31,120,180}
\definecolor{blind_safe_three_scheme_four_colors}{RGB}{178,223,138}
\definecolor{blind_safe_four_scheme_four_colors}{RGB}{51,160,44}

\definecolor{blind_safe_one_scheme_five_colors}{RGB}{240,249,232}
\definecolor{blind_safe_two_scheme_five_colors}{RGB}{186,228,188}
\definecolor{blind_safe_three_scheme_five_colors}{RGB}{123,204,196}
\definecolor{blind_safe_four_scheme_five_colors}{RGB}{67,162,202}
\definecolor{blind_safe_five_scheme_five_colors}{RGB}{8,104,172}

\definecolor{blind_safe_one_scheme_eight_colors}{RGB}{247,252,240}
\definecolor{blind_safe_two_scheme_eight_colors}{RGB}{224,243,219
}
\definecolor{blind_safe_three_scheme_eight_colors}{RGB}{204,235,197}
\definecolor{blind_safe_four_scheme_eight_colors}{RGB}{168,221,181}
\definecolor{blind_safe_five_scheme_eight_colors}{RGB}{123,204,196}
\definecolor{blind_safe_six_scheme_eight_colors}{RGB}{78,179,211}
\definecolor{blind_safe_seven_scheme_eight_colors}{RGB}{43,140,190}
\definecolor{blind_safe_eight_scheme_eight_colors}{RGB}{8,88,158}

\definecolor{blind_safe_one_scheme_seven_colors}{RGB}{118,42,131}
\definecolor{blind_safe_two_scheme_seven_colors}{RGB}{175,141,195}
\definecolor{blind_safe_three_scheme_seven_colors}{RGB}{231,212,232}
\definecolor{blind_safe_four_scheme_seven_colors}{RGB}{247,247,247}
\definecolor{blind_safe_five_scheme_seven_colors}{RGB}{217,240,211}
\definecolor{blind_safe_six_scheme_seven_colors}{RGB}{127,191,123}
\definecolor{blind_safe_seven_scheme_seven_colors}{RGB}{27,120,55}

\definecolor{yellow_one}{RGB}{255,255,212}
\definecolor{yellow_two}{RGB}{254,217,142}
\definecolor{yellow_three}{RGB}{254,153,41}
\definecolor{yellow_four}{RGB}{217,95,14}
\definecolor{yellow_five}{RGB}{153,52,4}

\definecolor{forestgreen}{RGB}{34, 139, 34}

\definecolor{hpca_one_of_ten_colors}{RGB}{158,1,66}
\definecolor{hpca_two_of_ten_colors}{RGB}{213,62,79}
\definecolor{hpca_three_of_ten_colors}{RGB}{244,109,67}
\definecolor{hpca_four_of_ten_colors}{RGB}{253,174,97}
\definecolor{hpca_five_of_ten_colors}{RGB}{254,224,139}
\definecolor{hpca_six_of_ten_colors}{RGB}{230,245,152}
\definecolor{hpca_seven_of_ten_colors}{RGB}{171,221,164}
\definecolor{hpca_eight_of_ten_colors}{RGB}{102,194,165}
\definecolor{hpca_nine_of_ten_colors}{RGB}{50,136,189}
\definecolor{hpca_ten_of_ten_colors}{RGB}{94,79,162}

\definecolor{MyLightBlue}{RGB}{141,211,199}
\definecolor{MyYellow}{RGB}{255,255,179}
\definecolor{MyLightPurple}{RGB}{190,186,218}
\definecolor{MyRed}{RGB}{251,128,114}
\definecolor{MyDarkBlue}{RGB}{128,177,211}
\definecolor{MyOrange}{RGB}{253,180,98}
\definecolor{MyGreen}{RGB}{179,222,105}
\definecolor{MyPink}{RGB}{252,205,229}
\definecolor{MyGray}{RGB}{217,217,217}
\definecolor{MyDarkPurple}{RGB}{188,128,189}

\colorlet{revcolor}{blue}

\usepackage{comment}
\newboolean{showcomments}
\setboolean{showcomments}{true}
\ifthenelse{\boolean{showcomments}}
{ \newcommand{\mynote}[3]{
     \fbox{\bfseries\sffamily\scriptsize#1}
        {\small$\blacktriangleright$\textsf{\emph{\color{#3}{#2}}}$\blacktriangleleft$}}}
{ \newcommand{\mynote}[3]{}}
\definecolor{asparagus}{rgb}{0.53, 0.66, 0.42}

\newcommand{\VariableStep}{$k$}

\AtBeginDocument{%
  }

\setcopyright{none} 
\copyrightyear{2024}
\acmYear{2024}
\acmDOI{XXXXXXX.XXXXXXX}

\acmISBN{978-1-4503-XXXX-X/18/06}

\begin{document}

\title{Count2Multiply: Reliable In-Memory High-Radix Counting}
%\subtitle{\normalsize{MICRO 2025 Submission
%    \textbf{\#913} -- Confidential Draft -- Do NOT Distribute!!}}
%\author{\normalsize{ISCA 2025 Submission
 %   \textbf{\#NaN} -- Confidential Draft -- Do NOT Distribute!!}}
\author{João P. C. de Lima$^{1,2}$, Benjamin F. Morris III$^{3}$, Asif Ali Khan$^{1}$, Jeronimo Castrillon$^{1, 2, 4}$, Alex K. Jones$^{5}$}
\affiliation{TU Dresden$^{1}$, ScaDS.AI$^{2}$, Barkhausen Institut$^{4}$ \country{Germany} }
\affiliation{Duke University$^{3}$, Syracuse University$^{5}$  \country{US} }
\email{Email(s): {joao.lima@tu-dresden.de, %ben.morris@duke.edu,asif\_ali.khan@tu-dresden.de,jeronimo.castrillon@tu-dresden.de,
akj@syr.edu}}


\begin{abstract}

Computing-in-memory (CIM) has been demonstrated across various memory technologies, ranging from memristive crossbars performing analog dot-product computations to large-scale digital bitwise operations in commodity DRAM and other proposed non-volative memory technologies. 
However, current CIM solutions face latency and reliability challenges. 
CIM fidelity lags considerably behind access fidelity.  Furthermore, bulk-bitwise CIM, although highly parallelized, requires long latency for operations like multiplication and addition, due to their bit-serial computation. 

This paper presents Count2Multiply, a technology-agnostic digital CIM approach to perform multiplication, addition and other operations using high-radix, massively parallel counting enabled by CIM bulk-bitwise logic operations. Designed to meet fault tolerance requirements, Count2Multiply integrates traditional row-wise error correction codes, such as Hamming and BCH, to address the high error rates in existing CIM designs. 
We demonstrate Count2Multiply with a detailed application to CIM in conventional DRAM due to its ubiquity and high endurance.  However, we note that the Count2Multiply architecture is compatible with other functionally complete CIM proposals. 
Compared to the state-of-the-art in-DRAM CIM method, Count2Multiply achieves up to $10\times$ speedup, $8\times$ higher GOPS/Watt, and $9.5\times$ higher GOPS/area, while outperforming GPU for vector-matrix multiplications.

\end{abstract}

\maketitle

\sloppy
% \vspace{-10pt}

\vspace{-.1in}
\section{Introduction}
\label{sec:intro}

Multiply and accumulate (MAC) is a fundamental computational primitive in many data-intensive application domains, including high-performance computing, machine learning, and bioinformatics. GPUs, TPUs, FPGAs, and other accelerators address these applications' needs with parallel execution units and/or integrated specialized MAC units. Despite delivering PetaFLOPs-scale performance, these architectures have substantial energy requirements and remain memory bound due to their compute-centric nature~\cite{boroumand2018google}. 
Consequently, there is a growing trend towards compute-in-memory (CIM) solutions~\cite{memristors-nn}. CIM has gained particular attention because emerging workloads often require only low-precision integer-integer ($\leq$ 8 bits), integer-binary, or integer-ternary operations for sufficient accuracy~\cite{li2016ternary,li2020rtn,ma2024era,wang2021bi}.

CIM solutions fully exploit the internal bandwidth and parallelism offered by memory arrays. These solutions are generally divided into two types. \textbf{Analog CIM} exploits current- or charge-sharing within memory arrays to compute the weighted sum between an input voltage vector and a column of memory cells.  This enables vector-matrix multiplication in constant time. 
However, it is mainly suitable for applications that can tolerate some loss in accuracy. Conversely, \textbf{digital CIM}, \textit{the focus of this paper}, performs computations more precisely, making it useful for a wider range of applications. In digital CIM, bulk-bitwise operations are commonly used across various memory technologies such as DRAM, SRAM, and emerging non-volatile memories (NVMs). These operations enable efficient execution of basic logic gates like \texttt{AND}, \texttt{OR}, and \texttt{NOT}, and serve as building blocks for more complex functions such as addition and multiplication. Early proposals for bulk-bitwise logic in DRAM and NVM designs emerged about a decade ago~\cite{magic, hamdioui2015memristor, pinatubo, seshadri2015fast}. However, it is only in recent years that experimental studies have shown real CIM capabilities in DRAM~\cite{COMP-DRAM, COTS-DRAM, yuksel2023pulsar} and emerging NVMs such as magnetic~\cite{lv2024experimental} and resistive~\cite{hoffer2020experimental,padberg2023experimental,brackmann2024experimental} memories.

However, CIM solutions face two key challenges to performance and accuracy, respectively, limiting their broader adoption:

\noindent\underline{Challenge 1}. 
Bulk bitwise operations are carried out in a bit-serial, word-/row-parallel fashion.  While this CIM approach offers higher throughput compared to traditional CPU and GPU systems~\cite{SIMDRAM,oliveira2025proteus} and also significantly reduces energy consumption, operation latency is high.  
Even with massive SIMD-style parallelism, control flow of bit-serial operations relies on \emph{sequential ripple propagation} such as in ripple-carry adders and multipliers. 
For instance, in Ambit~\cite{ambit}, each operation takes \SI{49}{\nano\s} compared to $<$\SI{1}{\nano\s} in CMOS.  This generally increases compute latency due to each operation's latency and the bit-serial steps in a full MAC operation. 

\noindent\underline{Challenge 2}. Fault rates for DRAM-based CIM range from $10^{-1}$ (experimental demonstration)~\cite{COTS-DRAM} to $10^{-6}$ (simulations)~\cite{elp2im}. Similar fault rates are observed in RRAM-based bitwise operations, ranging between $10^{-5}$ and $10^{-6}$ in experimental evaluations~\cite{brackmann2024experimental}.  This higher fault rate arises from the bitline current- or charge-sharing to the sensing circuitry resulting in reduced sense margins coupled with the impact of process variations~\cite{ambit, brackmann2024experimental}.   
Unlike memory accesses for which efficient error correcting codes have been developed, native error correction codes for CIM operations remain unsolved.  Moreover, memory error correction codes cannot be used directly as they are not homomorphic over CIM operations.

We make three observations to address these challenges:

\noindent\underline{Observation 1}.  Given the number of operations for bit-serial arithmetic is based on \textit{the precision of the operands}, tuning each MAC operation to the \textit{bit-width of each data element} can reduce the latency of CIM arithmetic.

\noindent\underline{Observation 2}.  Given that the latency of bit-serial arithmetic is based on \textit{the number of carry operations}, approaches like \textit{larger radices}, which reduce the number of carries, can reduce the latency of CIM arithmetic.

\noindent\underline{Observation 3}.  For any bitwise operation, there exists a sequence of additional bitwise operations that results in \texttt{XOR} or \texttt{XNOR}.  As traditional memory ECC codes are homomorphic over \texttt{XOR} or \texttt{XNOR}, these sequences can be checked using traditional syndrome checks.  

\begin{figure}[tbp]
\includegraphics[width=\columnwidth]{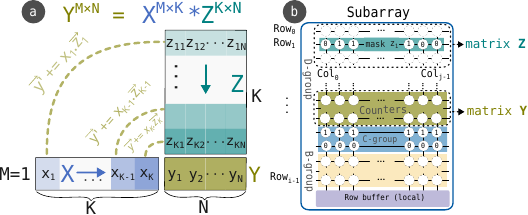}
\vspace{-0.10in}
\caption{Count2Multiply overview (a) integer vector, binary matrix multiplication example (b) DRAM subarray with counters and masks mapping, and Ambit's rows groups, i.e., computing (\emph{B-group}), control (\emph{C-group}) and data (\emph{D-group}).} %: (a) module, (b) chip, (c) subarray and (d) cell}
\label{fig:c2m-arch}
\vspace{-.25in}
\end{figure}

In this paper, we introduce Count2Multiply with a goal to simultaneously improve the performance including latency and throughput of CIM tensor computation.  Count2Multiply is a CIM architecture designed for integer vector/matrix ($\mathbb{X}$) binary matrix ($\mathbb{Z}$) computation illustrated in Fig.~\ref{fig:c2m-arch}a.  Instead of storing both tensor operators in memory as in prior work~\cite{pim-dram,SIMDRAM,ambit,elp2im}, we store $\mathbb{Z}$ packed-in memory rows to serve as counting masks.  The figure depicts this for integer vector $\mathbb{X}$.  The integer vector values $X_i$ are converted into in-memory processing commands, broadcast by the memory controller, to accumulate the values $X_i$ in high-radix counters stored column-wise in the memory.  Each column counter only adds $X_i$ to the counter if the bit from $\mathbb{Z}$ (e.g., $Z_{i,j}$) is `1'.  Count2Multiply is extensible to integer/vector, integer matrix operations through bit slicing $\mathbb{Z}$.  The Count2Multiply approach takes advantage of \underline{Observation 1} by tuning the number of CIM commands to the value $X_i$. 

This also takes advantage of \underline{Observation 2} by using high radix counters to minimize the number of carry operations required.  
We revisit classic circuit concepts to implement these counters with Johnson counters.  Both observations can be used together to further minimize count operations through this architecture using an early termination mechanism that significantly reduces latency and energy compared to existing CIM binary addition methods. 
Count2Multiply also considers \textit{reliability as a first-class optimization metric}.  Using \underline{Observation 3}, we explore the CIM operations to manage advancing the Johnson counters and show how they can still protect against individual bitflips even if they occur in different steps of calculating \texttt{XOR}.  Most current CIM designs lack built-in fault tolerance, relying mainly on replication and voting schemes~\cite{CORUSCANT,yuksel2023pulsar}.  Count2Multiply reintroduces fault-tolerance with traditional ECC codes to improve performance, storage overhead and fault tolerance over triple modular redundancy (TMR).
Building on the Ambit~\cite{ambit} CIM model, we demonstrate how Count2Multiply implements integer-binary and integer-integer matrix multiplications, and other operations from the machine learning and bioinformatics domains. 

Our concrete contributions are as follows:

\begin{itemize}
% \vspace{-.05in}
    \item We propose a high-radix in-memory counting methodology with optimizations that significantly enhance the performance of the MAC primitive in DRAM (Sec.~\ref{sec:CIMCounters}).  
    \item We illustrate how our in-memory counting mechanism can be used to execute massively parallel integer-binary matrix multiplication, also extended to integer matrix-matrix multiplications through bit-slicing, addition and other operations (Sec.~\ref{count2multiply:presentation}).
    \item We demonstrate how traditional row-level ECC can be leveraged to protect CIM operations in Count2Multiply, while still protecting row-level accesses (Sec.~\ref{sec:fault-tolerance}). 
    \item We evaluate Count2Multiply on multiple applications from the bioinformatics and machine learning domains and compare performance to state-of-the-art in-DRAM designs and high-end GPUs (Sec.~\ref{sec:eval}). 
\end{itemize}

Count2Multiply is a CIM architecture, which is technology-agnostic and applicable to any functional complete bulk-bitwise CIM proposal, including other DRAM implementations~\cite{COTS-DRAM} and NVM CIM implementations~\cite{magic,hamdioui2015memristor,pinatubo,seshadri2015fast,lv2024experimental,hoffer2020experimental,padberg2023experimental,brackmann2024experimental}. 
Compared to prior in-DRAM designs~\cite{SIMDRAM}, our proposed design and optimizations improve execution time by up to $10\times$, delivering $8\times$ higher GOPS/Watt, and $9.5\times$ higher GOPS/area, on average. 
\section{Background and Related Work}
\label{sec:related}
This section provides background on DRAM, including its CIM capabilities, Johnson counters, CIM faults and fault tolerance schemes. 

\vspace{-0.08in}
\subsection{DRAM Organization and Operation} 
\label{subsec:dram-CIM}

\begin{figure}[tbp]

\includegraphics[]{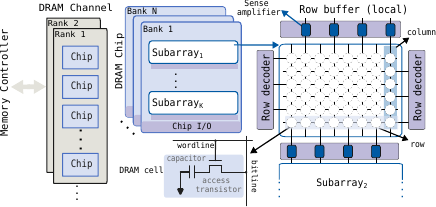}

\caption{DRAM organization} 
\label{fig:dram}

\end{figure}

Fig.~\ref{fig:dram} illustrates the hierarchical organization of a modern DRAM system. 
A CPU manages memory through multiple memory controllers (one per channel), each handling memory read, write, and refresh operations. Each channel connects to one or more DRAM modules, which contain ranks composed of multiple DRAM chips working in lockstep. 
A DRAM chip contains multiple banks that share an internal bus, connecting them to the chip’s I/O circuitry. Each bank consists of several subarrays (e.g., 16-32), each with its peripheral circuitry (e.g., row decoders, sense amplifiers, and local wordline drivers) for data manipulation. 
Within a subarray, a DRAM row shares a wordline, which is activated by a row decoder to enable read or write operations.
A DRAM cell, consisting of a capacitor and a transistor, is organized into a 2D array of rows and columns, each subarray connecting to a row of sense amplifiers (SA) also known as a local row buffer. 

Reading or writing data to a DRAM row is a three-step process. First, the row is \emph{activated} (\texttt{ACT}), bringing cell data to the row buffer (e.g., sense amplifiers). Second, the \emph{read/write} (\texttt{RD}/\texttt{WR}) operation transfers the data from/to the row buffer to/from the bus. Third, the row is \emph{precharged} (\texttt{PRE}), restoring bitlines to a stable state for the next operation.
The memory controller schedules commands regulated by a set of timing parameters that ensure sufficient delays between commands to correctly retrieve and retain data in DRAM cells. For example, \texttt{tRAS} (Row Active Time) defines the minimum time a row must remain active before it can be precharged, ensuring data is fully accessible.

\vspace{-.1in}
\subsection{Compute-In-DRAM}
Previous research has shown that certain functions can be performed directly \emph{in memory} by carefully modifying the standard sequence of DRAM operations~\cite{rowclone,ambit,COMP-DRAM}. 
For instance, RowClone (\texttt{RC})~\cite{rowclone} copies \texttt{src} row to \texttt{dst} row within the same subarray using back-to-back \texttt{ACT} commands followed by a \texttt{PRE} command, known as \emph{activate-activate-precharge} (\texttt{AAP}). 
The \texttt{AAP} sequence operates by first activating the \texttt{src} row to drive its contents onto the bitlines. Activating the \texttt{dst} row then transfers these values to overdrive its capacitors~\cite{rowclone}. Finally, a precharge command resets the subarray for the next operation.
In addition to RowClone, DRAM implements logic operations using \textit{multirow activation} (MRA), wherein multiple rows in a subarray are activated \textit{simultaneously}, followed by a \texttt{PRE} command, known as \emph{activate-precharge} (\texttt{AP}). 
In-DRAM CIM designs achieve simultaneous MRA either through a custom row decoder~\cite{ambit, SIMDRAM} or by violating memory timings to issue consecutive \texttt{ACT} commands~\cite{COMP-DRAM, COTS-DRAM, elp2im}.

Ambit~\cite{ambit} uses triple-row activation (\texttt{TRA}) to perform a bulk bitwise majority (\texttt{MAJ3}) function, where a bitline reflects the majority state of three cells. For functional completeness, \texttt{NOT} is implemented using dual-contact cells (DCCs), which consume two wordlines to connect a capacitor to either the bitline or $\overline{\text{bitline}}$ in the sense amplifier circuit.
To simplify row decoding, Ambit divides the space of row addresses in each subarray into three groups, as shown in Fig.~\ref{fig:c2m-arch}b: (i) B-group, eight rows for bulk bitwise \texttt{MAJ3}/\texttt{NOT}, (ii) C-group, two rows storing `0' (\texttt{C0}) and `1' (\texttt{C1})
, and (iii) D-group, the remaining $r-10$ rows for data storage, where $r$ is the total row count per subarray. 
Though the B-group contains only 8 rows, it is responsible for 16 unique addresses -- these addresses map to different combinations of 1, 2, or 3 rows -- enabling the row activations required for access, copying and computing \texttt{MAJ3}, respectively.

Commercial-off-the-shelf (COTS) DRAMs can perform boolean functions by executing carefully engineered sequences of DRAM commands that activate multiple rows, referred to as functionally complete DRAM (or FCDRAM)~\cite{COTS-DRAM}. 
The key command sequence, \emph{activate-precharge-activate} (\texttt{APA}), activates rows in \emph{neighboring subarrays} that share SAs. Leveraging initialization of rows to fractional values~\cite{gaofracdram}, \texttt{AND} and \texttt{OR} operations use two reference rows -- \texttt{Vdd} and \texttt{Vdd/2} for \texttt{AND} and \texttt{Gnd} and \texttt{Vdd/2} for \texttt{OR} -- in one subarray, while operand rows (\texttt{A}, \texttt{B}) reside in the other. 
FCDRAM claims \texttt{NOT} is obtained when the negated value of \texttt{src} row is written to \texttt{dst} row in the neighboring subarray\footnote{FCDRAM notes that \texttt{NOT} is limited to DRAMs constructed only with true cells like SK Hynix and Samsung~\cite{COTS-DRAM}.}.  
In this work, we require support for copying the inverted results from the neighboring subarray back to the original.
Both Ambit and FCDRAM multi-row operations in COTS-DRAM are destructive, as all activated rows are overwritten with final computation result by the end of the command sequence.

\vspace{-.1in}
\subsection{Fault Modes and Fault Tolerance for CIM}
\label{subsec:reliab-bg}

Violating DRAM timing parameters facilitates CIM but increases bit error rate~\cite{ambit,elp2im,COTS-DRAM, yuksel2023pulsar}.
In simulations, the interaction between activated rows and bitlines to compute \texttt{MAJ} works reliably under an idealized scenario, assuming DRAM cells have rather small variability (<6\% in~\cite{ambit}) and transistors and bitlines operate without deviation.
However, in practical implementations, process variation induces non-uniform electrical characteristics across cells, resulting in instability in multi-row activation leading to faults.

Only minimal work beyond TMR exists for CIM and all of it is for memories other than DRAM.  For computing using Spintronic Racetracks, a recent technique called CIRM-ECC protects transverse read-based logic operations~\cite{cirm-ecc}. 
A recent work in RRAM constructs the parity bits in specialized peripheral circuity following each logic level of the CIM computation but results in requires additional area overhead, and increases critical path latency~\cite{nvm-logic-cim-ecc}.  
In Sec.~\ref{sec:fault-tolerance}, we propose a reliability scheme that uses \textit{existing} ECC circuitry to protect memory access and CIM operations in Count2Multiply.  

\vspace{-.1in}
\subsection{Johnson Counters}
\label{subsec:JC}
Johnson counters (JC)~\cite{JC_patent}, also known as twisted ring counters, are cyclic shift register-based sequential circuits with single-bit transitions between consecutive states.  This feature also minimizes transitional errors.  For example, a 5-bit JC with the least significant to most significant bits moving left to right progresses through states as: $\texttt{10000}(1) \rightarrow \texttt{11000}(2)$ ...  $\rightarrow \texttt{11111}(5)$ $\rightarrow \texttt{01111}(6)$  ... $\rightarrow \texttt{00001}(9) \rightarrow \texttt{00000}(0)$, maintaining the cyclic property where decimal 9 rolls over to 0. This structure allows an $n$-bit counter to represent $2n$ distinct states.  
A JC counter increment shifts the register's content from the least significant bit 
(\emph{forward shift}) while inverting the most significant bit as a feedback bit to the new least significant bit (\emph{inverted feedback}).  Decrement would happen in reverse with a \textit{backward shift} followed by a \textit{inverted feed-forward}.

\begin{figure}[]
    \centering
    \subfloat[Short-read token repetition] {
    \includegraphics[width=0.47\columnwidth]{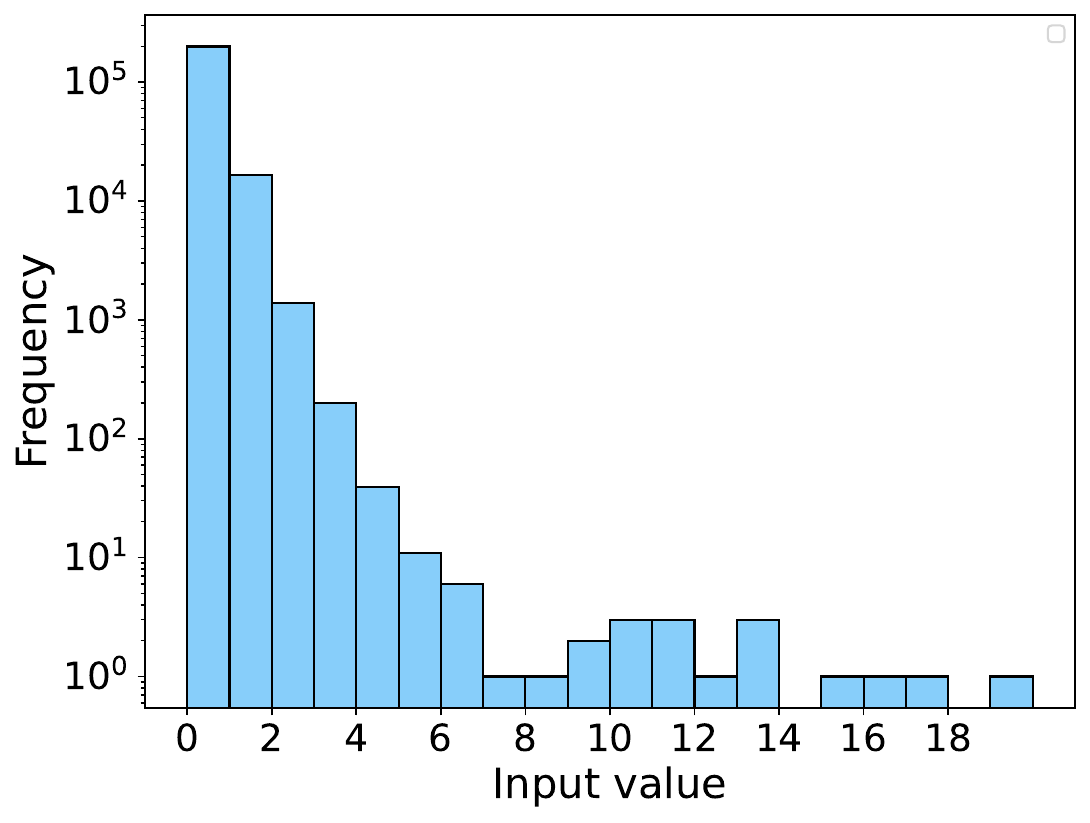}
    \label{subfig:dna_input}
    }
    \subfloat[8-bit input embeddings] {
    \includegraphics[width=0.47\columnwidth]{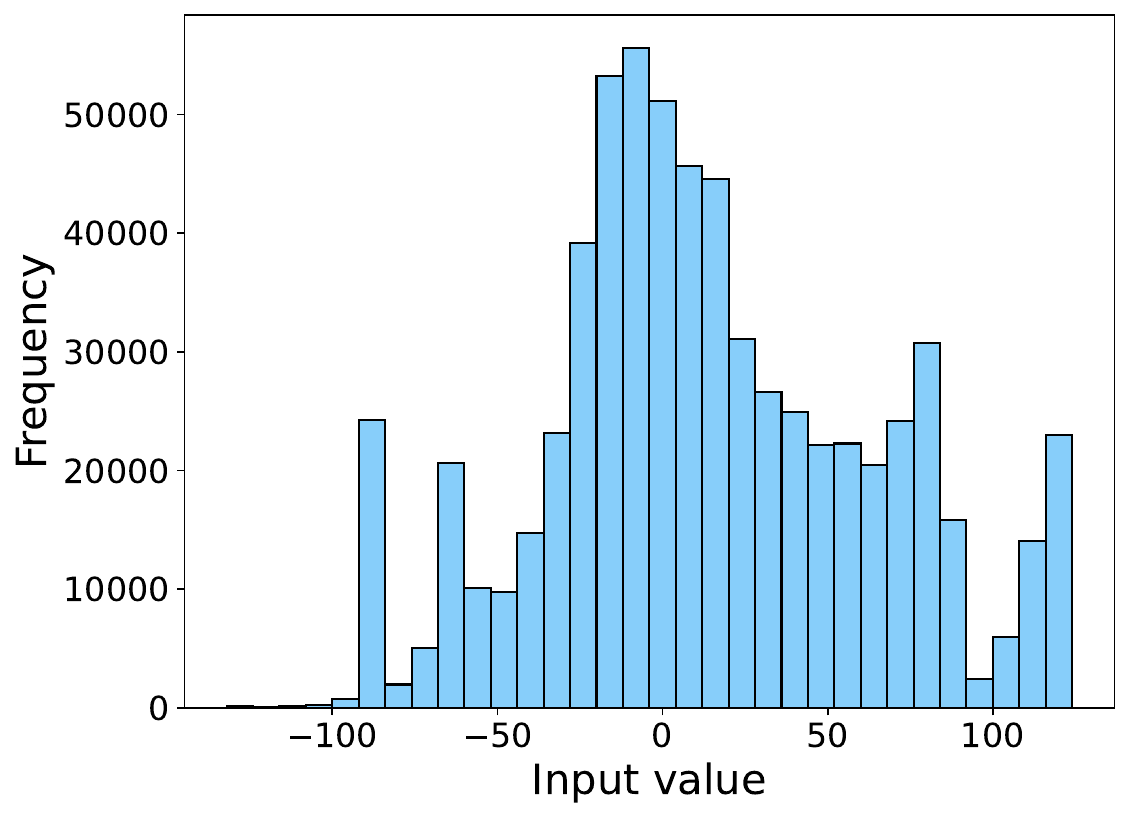}
    \label{subfig:bert_input}
    }
    \vspace{-0.12in}
    \caption{Input distribution in DNA pre-alignment filtering and BERT language model.  Values are small (circa 4--8 bits).}
    \label{fig:histogram}
     \vspace{-.2in}
\end{figure}

%\subsection{Rationale of High-Radix Addition}
\vspace{-.05in}
\section{Motivation} % for High-Radix Addition}
\label{subsec:motiv}

Several recent works on low-precision matrix multiplication have attempted to reduce the cost of MAC by using bitslicing~\cite{you2020shiftaddnet, CORUSCANT, dracc, chopper} or using Ternary Weight Networks (TWNs)~\cite{chee2024quip, ma2024era}. With TWNs, multiplications are replaced by \emph{masked additions} for more efficient CIM implementations~\cite{dracc, pim-dram}. 
However, CIM arithmetic generally requires bit-serial addition in the form of ripple carry adders (RCAs).  
To maximize the performance, this approach leverages element-parallel (e.g., vector-style) computing to add many values simultaneously~\cite{pim-dram,leitersdorf2023aritpim,SIMDRAM}. Yet, additions in this computation style suffer from long carry propagation chains, even when many smaller values are added to a larger sum. Worse, managing these long carry chains can be made entirely \emph{unnecessary}; many applications from different domains feature accumulation of narrow-range values, as demonstrated in Fig.~\ref{fig:histogram} where in DNA pre-alignment filtering (\ref{subfig:dna_input}) and BERT (\ref{subfig:bert_input}), the values for accumulation are representable in 4--8 bits. For every new addition of these smaller values, RCA-based accumulation must often fully process the carry propagation of larger values (circa 32 bits) due to a large accumulated total. 

\begin{figure}
    \centering
    \subfloat[Accumulated error of adds]{
    \includegraphics[width=0.47\columnwidth]{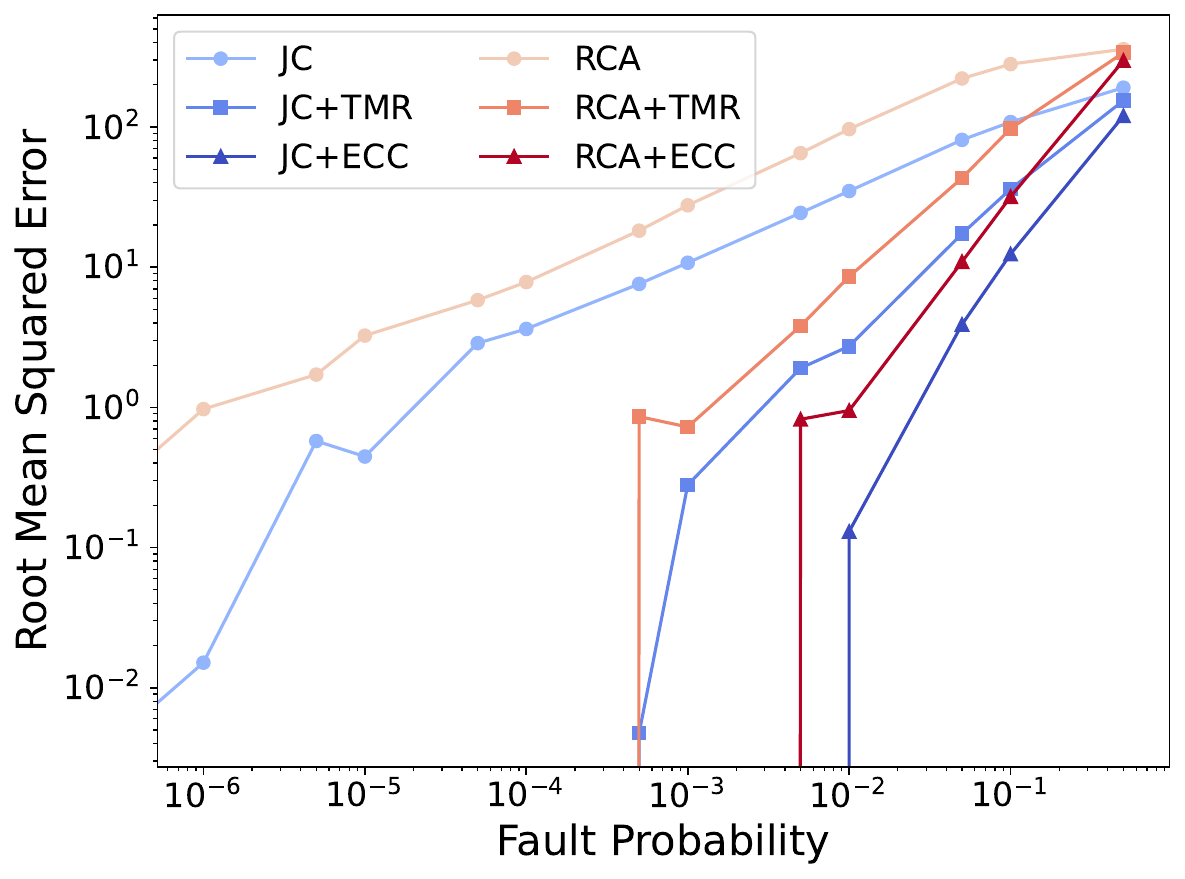}\label{subfig:basic_add}}
    \hfill
    \subfloat[Fault impact on DNA filtering]{
    \includegraphics[width=0.47\columnwidth]{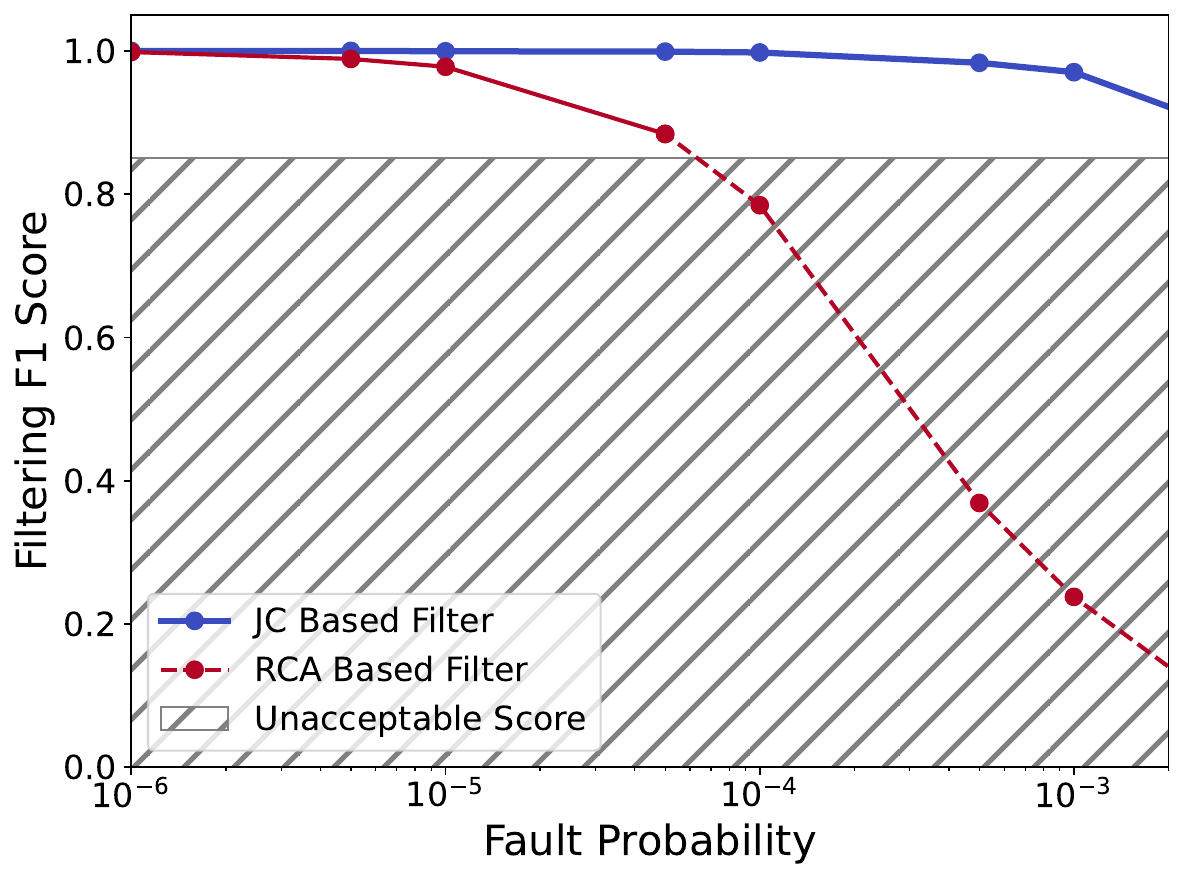}\label{subfig:dna_motiv}}
    \vspace{-0.12in}
    \caption{Fault rate impact on application accuracy. }
    \label{fig:motiv_Add}
    \vspace{-0.17in}
\end{figure}

The additional CIM operations from carry chain management not only increase the latency but also substantially impact reliability as more faulty CIM operations have the potential to perturb higher order bits of the accumulated value. 
We compare the fault rate and impact of accumulated faults for RCA-based and radix-10 JC implementations to implement DNA prealignment filtering in Fig.~\ref{fig:motiv_Add}.  Comparing Root Mean Squared Error (RMSE) (\ref{subfig:basic_add}), RCA shows substantial error with an incident fault rate of $10^{-6}$, while JC can tolerate faults up to $10^{-5}$ to achieve the same error rate.  The impact of these errors cannot be entirely addressed by a fault tolerant algorithm; the filtering score of near unity with low RMSE drops quickly when errors are introduced, indicated by the gray region in Fig.~\ref{subfig:dna_motiv}. The JC-based filter staves this off into much higher fault rates than the RCA-based filter.

Notably, a JC implementation is more fault-tolerant than using RCAs regardless of the error correction approach. We also note that redundancy and voting are inherently inefficient: TMR has a circa $4\times$ overhead in operation count (three repeated operations and the voting operation) for both encodings (RCA+TMR and JC+TMR). In addition, TMR has a higher error rate than single error correction codes (RCA+ECC and JC+ECC).   This motivates our Count2Multiply approach that leverages JC accumulation and introduces integrated error correction based on traditional ECC codes rather than replication and voting. In the next section, we describe how JC operations are realized in-memory.  

\newcommand{\VariableCounters}{C} 
\section{In-Memory High-Radix Counters}
\begin{figure}[tbp]
\centering
\subfloat[initial] {
\includegraphics[height=1.5in]{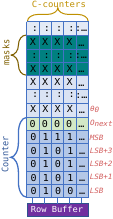}
\label{subfig:counter-a}
}\hfill
\subfloat[all count] {
\includegraphics[height=1.5in]{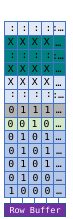}
\label{subfig:counter-b}
}\hfill
\subfloat[masked] {
\includegraphics[height=1.5in]{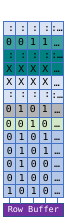}
\label{subfig:counter-c}
}\hfill
\subfloat[multi-digit] {
\includegraphics[height=1.5in]{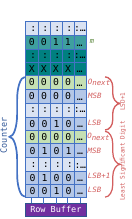}
\label{subfig:counter-d}
}\hfill
\vspace{-0.1in}
\caption{$\VariableCounters$, 5-bit JCs in memory: (a) before counting; (b) all counters count; (c) masked counting; (d) multi-digit counting.}
\label{fig:DRAMCounter}
\vspace{-0.17in}
\end{figure}

%\subsection{Basic Counter Design}
\label{sec:CIMCounters}
To realize the Count2Multiply architecture, we first discuss our in-memory high-radix counter implementation.  High-radix counters are built from multi-bit ``digits'' at the specified radix and can consist of sufficient digits to store a determined maximum value for the computation.  To implement these counters in memory, we define a counter as dedicated memory rows shown in Fig.~\ref{subfig:counter-a}, such that all bits of a counter reside in the same column. Specifically, we adopt Johnson Counters (JC) (see Sec.~\ref{subsec:JC}) as it requires fewer operations than an RCA for a single increment.  Thus, we can partition the rows to represent individual digits encoded as JCs.
Let us consider a single-digit $n$-bit JC that can count from 0 to $2n-1$, i.e., radix-$2n$. For instance, $n=5$ corresponds to a single-digit base-10 counter, from 0 (``\texttt{00000}'') to 9 (``\texttt{00001}''). 
Our $n$-bit JC requires $n+1$ memory rows (one row for overflow, $O_{next}$). 
In the following sections, we outline how to increment JCs in parallel with CIM operations, handle overflows and operate on multi-digit counters.

\vspace{-.1in}
\subsection{Single-Digit Unit Increment}
\label{sec:single-digit-unit}
Fig.~\ref{subfig:counter-a} illustrates an example of $\VariableCounters$ JCs with their bit positions---MSBs and LSBs---labeled on the right. 
To increment all $\VariableCounters$ counters, all bit positions are \emph{forward shifted} by one towards the MSB, while the LSB is computed by taking the \emph{inverted feedback} of the MSB (LSB~$\gets~\overline{\text{MSB}}$). 
To achieve this with CIM, the MSB row is initially cloned into a temporary row in the array, e.g., $\theta_0$.  
Subsequently, each bit row, except the MSB, is moved---i.e., ``shifted''---by one position towards higher significance using \texttt{RC}~\cite{rowclone} operations, implementing the \emph{forward shift} steps. Finally, $\theta_0$ is inverted and cloned to the LSB row using a \texttt{NOT} operation (see Sec.~\ref{subsec:dram-CIM}), hence implementing the \emph{inverted feedback}. 
Fig.~\ref{subfig:counter-b} shows the final result of all $\VariableCounters{}$ 5-bit JCs after a unit increment (requiring four \emph{forward shift} and one \emph{inverted feedback} steps).

\vspace{-.1in}
\subsection{Single-Digit Masked Unit Increment}
\label{subsec:masked-counting}
Some applications, particularly those involving tensor operations, require selective increments or \emph{masked counting}, where only a subset of counters is updated based on a stored mask $m$. This mask is stored in a row within the subarray containing the counters (highlighted row in Fig.~\ref{subfig:counter-c}). 
The masked \emph{forward shift} and \emph{inverted feedback} are implemented using the following logical expressions:   
\[
\begin{aligned}
b_i &= (b_{i} \land \overline{m}) \lor (b_{i-1} \land m), \quad \text{where } i \in \{n, n-1, \ldots, 2\} \\
b_1 &= (b_1 \land \overline{m}) \lor (\overline{b_n} \land m)
\end{aligned}
\]
where $b_i$ represents the counter bit at index $i \in \{\text{LSB}...\text{MSB}\}$.

\begin{figure}
    \centering
    \centering
    \subfloat[MIG-based circuits]{
    \includegraphics[height=1.5in]{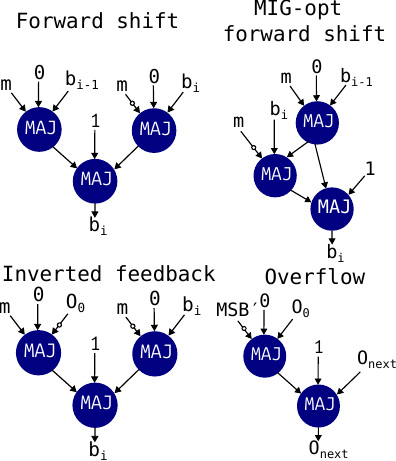}\label{subfig:maj_circuits}}
    \hfill
    \subfloat[$\mu$Program for Ambit~\cite{ambit}]{
    \includegraphics[height=1.5in]{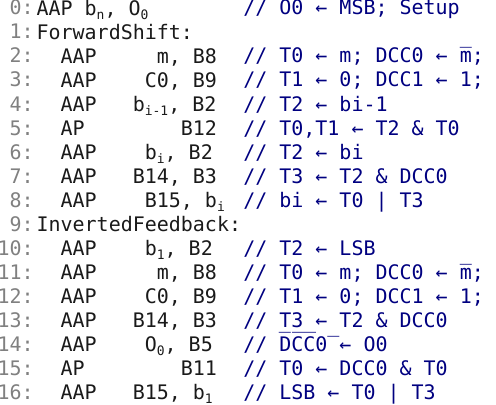}\label{subfig:microProgram}}
     \vspace{-0.1in}
    \caption{Majority-based in-DRAM operations for counting and overflow detection.\protect\footnotemark}%\footnotemark{}.}
    \label{fig:cim-ops}
    \vspace{-0.2in}
\end{figure}

For every bit position ($b_i$) in the forward shift, the increment process requires two \texttt{AND}, one \texttt{OR}, and one \texttt{NOT} operation. The inverted feedback requires an additional \texttt{NOT} operation to invert the MSB. Fig.~\ref{fig:cim-ops} shows the Majority-Inverter Graphs (MIGs)~\cite{amaru2014majority} and $\mu$Program for both forward shift and inverted feedback using Ambit's~\cite{ambit} primitives.
As illustrated, we construct \texttt{AND} and \texttt{OR} from the \texttt{MAJ3} function. We first synthesize this expression into a MIG (Fig.~\ref{subfig:maj_circuits}), and subsequently employ MIG-based optimizations, similar to prior works~\cite{amaru2014majority,EPFLLibraries}.  This minimizes the number of \texttt{RC}s in both the forward and backward shifts by scheduling the \texttt{MAJ3} operations and allocating the rows in the B-group to maximize data reuse and reduce the number of row initialize operations, i.e., cloning constant `0' or `1' control rows for \texttt{AND} or \texttt{OR} to a compute row. 

Fig.~\ref{subfig:microProgram} presents the optimized sequence of seven \texttt{AP}/\texttt{AAP} memory commands for each step of the counter increment. Using \texttt{AAP} operations, the mask row $m$, constant zero \texttt{C0}, and the counter bit $b_{i-1}$ are cloned to three rows in the B-group (Line 2--4 in Fig.~\ref{subfig:microProgram}), followed by \texttt{MAJ3} to implement \texttt{AND} (Lines 5,7) and \texttt{OR} (Line 8) operations. The \texttt{NOT} operation is inherently handled by a \texttt{RC} operation with special DCC rows~\cite{ambit}, so obtaining $\overline{m}$ incurs no additional overhead. This process repeats for all $n$ bits in the counter ($n-1$ forward shift and 1 inverted feedback step), for a total of $7n$ operations.

\begin{comment}
    \centering
    %\includegraphics[width=0.9\linewidth]{figures/dram-cim-ops.png}
    \includegraphics[width=\linewidth]{figures/circuits_microProgram.pdf}
    \caption{Majority-based in-DRAM operations for counting and overflow detection.\protect\footnotemark}%\footnotemark{}.}
    \label{fig:cim-ops}
    \vspace{-0.2in}
\end{comment}

\footnotetext{We modified Ambit's B-group mapping~\cite{ambit} so address \texttt{B11} activates \texttt{T0}, \texttt{T1}, and \texttt{DCC0}. This does not affect prior operations as \texttt{B11} was unused.}

\vspace{-.1in}
\subsection{Overflow Detection in Single-Digit Counters}
\label{subsec:overflow}
For single-digit counters (Fig.~\ref{subfig:counter-a}-\ref{subfig:counter-c}),
when a counter rolls over (\emph{overflows}), this information is retained in a dedicated row ($O_{next}$). An overflow is detected when the JC's MSB transitions from `1' to `0'. 
In masked counting, each increment can potentially cause an overflow in one or more of the $\VariableCounters$ counters. The \emph{overflow detection} is expressed as: $O_{next} \leftarrow O_{next}$ \texttt{OR} $\theta_0$ \texttt{AND} $\overline{\text{MSB}'}$, where $\text{MSB}'$ is the new MSB after shifting (\emph{forward shifts} and \emph{inverted feedback}).
Calculating $O_{next}$, as illustrated in Fig.~\ref{subfig:maj_circuits}, requires a total of six \texttt{AAP} operations (4 \texttt{RC} and 3 \texttt{MAJ3} operations).

\vspace{-.1in}
\subsection{Multi-Digit Increment} 
\label{subsec:multi-digit-count}
For $D$-digit counters (Fig.~\ref{subfig:counter-d}), all digits belonging to a counter are stored in the same column, requiring $D \cdot (n+1)$ rows. Starting from the least significant \emph{digit} (LSD), the counter digits are updated sequentially using the single-digit counting mechanism. 
After $2n$ increments (\emph{forward shift}, \emph{inverted feedback} and \emph{overflow detection}) to the $LSD$, $O_{next}$ is used to increment the $\text{LSD}+1$ and subsequent digits through $O_{next}$ rippling. 
Note, so far we have only discussed single-digit inputs. However, in practice, inputs can also have multiple digits, with each digit ranging between 0 and $2n-1$. In such cases, for counter accumulation, i.e., adding the \emph{same multi-digit input} to all \emph{multi-digit counters in memory}, we align digits with equal significance and update all digits sequentially from the LSD, resolving carries as we move to the most significant digit (MSD).
The increment process is repeated $D +\sum_{i=1}^{D} d_i$ times, where $d_i$ is the digit value of the input in the base $2n$. The \emph{digit sum} represents the number of \emph{unit increments} triggered by the multi-digit input, while the $D$ addend accounts for the unit increments required for \emph{carry rippling}. This sequential digit increment remains costly, motivating a series of optimizations in the next section (Sec.~\ref{subsec:count-opt}) to reduce the number of CIM operations.

\noindent\textbf{Overflow Detection and Carry Rippling.} In multi-digit counters, when the current counter digit overflows, the next higher digit must be incremented. A detected overflow in the LSD generates a carry ($O_{next}$) to the next significant \emph{digit}, i.e., $\text{LSD}\rightarrow\text{LSD}+1$, which may ripple through to the MSD, depending on the counters state. 
A na{\"i}ve implementation of \emph{carry rippling} %(Fig.~\ref{fig:overflow_minimization}) 
would fully resolve detected overflows in the LSD, propagating carries through all digits before proceeding to the next increment.\footnote{\emph{Digit-wise carry ripple:} unit increment to the next higher digit using $O_{next}$ as a mask.} 
This is inefficient because an overflow check after every increment is unnecessary, especially for unit and small value increments (see Fig.~\ref{fig:histogram}).
To optimize this critical path operation, we use one dedicated $O_{next}$ row per digit to retain the pending overflow until it is unavoidable (see Sec.~\ref{sss:overflow_minimization}).

\noindent\textbf{Decrements.} For negative inputs, the counter is decremented through \emph{backward shifts} and \emph{inverted feed-forward}. The \emph{underflow} detection mechanism is similar to overflow except the MSB transitions from zero to one and the $O_{next}$ bit is used to decrement the next significant digit accordingly. Outstanding overflows or underflows must be resolved before switching from increment to decrement and vice versa, or a row representing a sign-bit of overflow must be allocated, $O_{sign}$.

\vspace{-.1in}
\subsection{Optimized Counter Design}
\label{subsec:count-opt}
Counters with \emph{unit increments} and \emph{digit-wise carry rippling} %in Ambit 
are more costly than directly implementing ripple-carry addition (Fig.~\ref{subfig:k-ary}). 
In this section we propose scheduling optimizations to significantly improve all counters' performance.

\subsubsection{Variable-Step (k-ary) Increment}
\label{sss:k-ary-inc}
This section presents an optimization to perform an increment by \VariableStep, where $1 \leq$ \VariableStep~$\leq 2n-1$, with the same number of steps as a unit increment; increment by one and increment by \VariableStep{} have the same latency. 
The cyclic property with a k-ary increment remains guaranteed, meaning the JC state rolls back to 0 when the sum exceeds the counter capacity ($2n-1$). 
For instance, a 5-bit JC can transition directly from $\texttt{10000}(1) \rightarrow \texttt{00111}(7)$ and $\texttt{00111}(7) \rightarrow \texttt{11100}(3)$ when incrementing \VariableStep~$=6$. Overflow detection indicates whether an increment results in an overflow (Sec.~\ref{subsec:overflow}).

\begin{figure}[tbp]
    \centering
    \includegraphics[width=\linewidth]{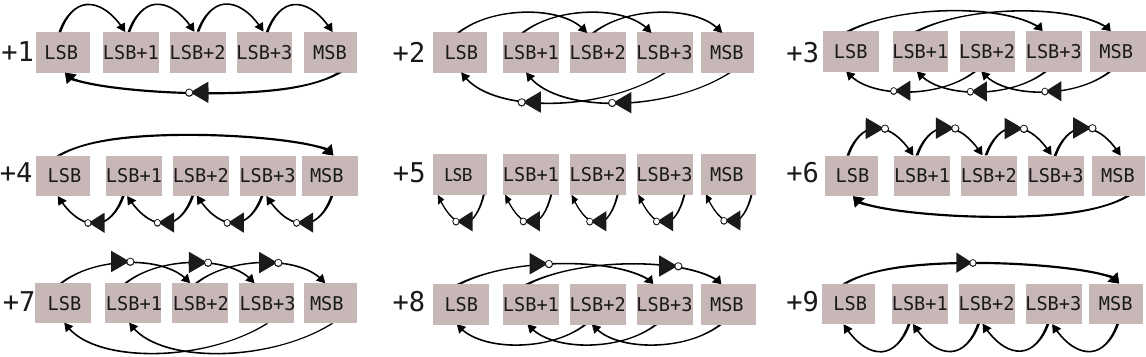}
    \vspace{-.2in}
    \caption{Transition patterns of a 5-bit counter (radix-10) for incrementing any value between 1 and 9.}
    \label{fig:k-ary}
    \vspace{-.15in}
\end{figure}

For any \VariableStep-ary increment, while the number of fundamental steps (\emph{forward shift} and \emph{inverted feedback}) remain the same as in unit increment, the shifting patterns are different. Fig.~\ref{fig:k-ary} illustrates all patterns for achieving a direct state transition from any radix-10 JC's value to its value incremented by \VariableStep.
For instance, an increment by \texttt{+2}, equivalent to applying the unit increment (\texttt{+1}) twice, is simplified into three \emph{forward shifts} and two \emph{inverted feedback} steps. The same principle applies to all patterns in Fig.~\ref{fig:k-ary}. These transition patterns correspond to \VariableStep ~distinct %$\mu$Programs
sequences of CIM commands, yet all require the same number of CIM (i.e., \texttt{AAP}) commands.

Algorithm~\ref{alg:johnson_counter_variable_step} generalizes the transition pattern generation, illustrating how each counter bit is updated depending on the value of \VariableStep ~with respect to $n$. The bit manipulation depends on whether the increment amount is less than or greater than $n$. If less than or equal to $n$, Lines 2-7 conditionally update the counter state using a mask $m$ and calculate the digit overflow accordingly. This process is illustrated by the increments from 1 to 5 in Fig.~\ref{fig:k-ary}. Line 3 corresponds to the upper arrows, while Line 5 represents the lower arrows with a \texttt{NOT} operation. Similarly, for increments from 6 to 9, Line 12 corresponds to the upper arrows, and Line 10 represents the arrows with a \texttt{NOT} operation. With \VariableStep-ary transitions, counting requires $2\cdot (7n+7)$ operations per \emph{input digit}, as each \VariableStep-ary increment may propagate a carry, leading to an additional \emph{carry rippling} command sequence. This occurs because \VariableStep-ary transitions do not take a carry input, making carry propagation a separate operation. In the following section, we present a method to minimize the cost of this cascading effect in counters. 

\begin{figure}[tbp]
    \centering
    \subfloat[Unit vs. $k$-ary increment] {
    \includegraphics[height=1.35in]{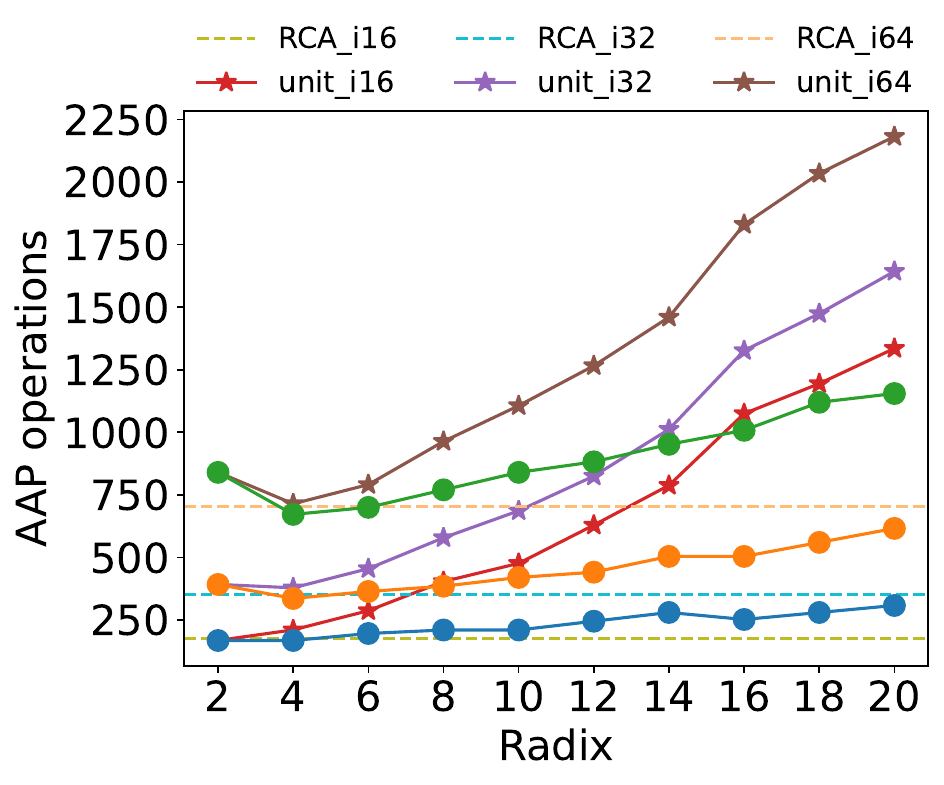}
    \label{subfig:k-ary} 
    }
    \subfloat[$k$-ary only vs. IARM] {
    \includegraphics[height=1.35in]{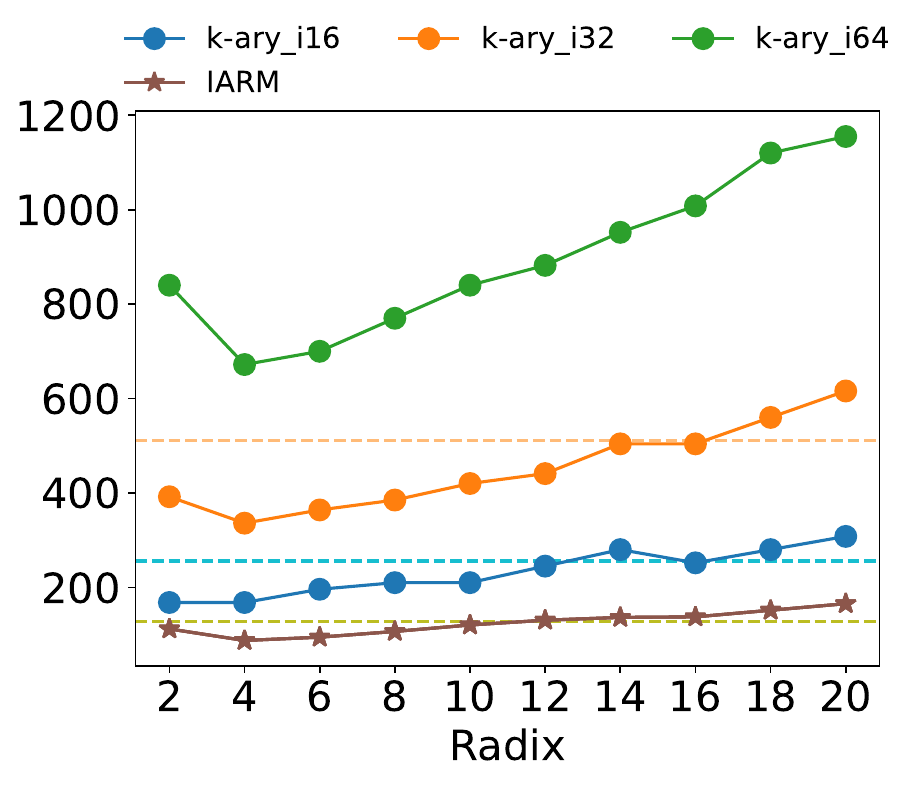}
    \label{subfig:OVM}
    }\hfill
\vspace{-.15in}    
    \caption{Masked addition performance for unit, k-ary, carry rippling minization (IARM), and ripple-carry adders (RCA).}
    \label{fig:performance}
    \vspace{-0.1in}
\end{figure}

Fig.~\ref{subfig:k-ary} compares the average number of \texttt{AAP} operations required for different bases when accumulating (uniform distribution) of 8-bit inputs on counters with capacities equivalent to 16, 32, and 64-bit \texttt{ints}.\footnote{A $D$-digit counter with $n$ bits per digit has a capacity of $(2n)^D$. We size counters to meet or exceed the capacity of binary integers by adding more digits.} \VariableStep-ary counting provides a 2--6$\times$ reduction in CIM operations over unary counting for varying-radix counters.

\begin{algorithm}[tbp]
%\footnotesize
\scriptsize
\caption{Variable-step increment for $n$-bit JC}
\label{alg:johnson_counter_variable_step}
\KwIn{Johnson counter $C \leftarrow [b_{0}, \dots , b_{n-1}]$, mask $m$, increment amount \VariableStep} 
%$R = \{ C_{n-j} \mid j \in [0, \VariableStep] \}$\;
 %\tcc{$b_1$ is LSB and $b_{n}$ is MSB}
 \uIf {\VariableStep~$\leq n$}{
 %\For{$i \leftarrow n-1$ \KwTo $k$ \KwDo $i \leftarrow i-1$}{

     %\For {each bit $b_i$ from $n-1$ to $\VariableStep$ step  $-1$}{
    
     %\For {$b_i$ from $\{ C_{n-j} \mid j \in [0, n-$\VariableStep$) \}$}
     \For{$i\gets n-1$ \textup{\textbf{ downto }} $k$}{
         $b'_i \leftarrow (\overline{m} \land b_i) \lor (m \land b_{i-k})$; \tcp{Forward Shift}
     }
     %\For {$b_i$ from $\{ C_{j} \mid j \in [1, $\VariableStep$] \}$}{
     \For{$i\gets 0$ \textup{\textbf{ upto }} $k$ }{
         $b'_i \leftarrow (\overline{m} \land b_i) \lor (m \land \overline{b_{n-k+i}})$; \tcp{Inverted Feedback}
     }
     $O_{\text{next}}' \leftarrow O_{\text{next}} \lor (b_{n-1} \land \overline{b_{n-1}'})$; \tcp{Overflow Checking} 
 }
 \Else{
     \VariableStep~$\leftarrow$~\VariableStep$- n$\;
    %\For {each bit $b_i$ from $n-1$ to $\VariableStep$ step $-1$}{
    %\For {$b_i$ from $\{ C_{n-j} \mid j \in [0, n-$\VariableStep$) \}$}{
    \For{$i\gets n-1$ \textup{\textbf{ downto }} $k$}{
         $b'_i \leftarrow (\overline{m} \land b_i) \lor (m \land \overline{b_{i-k}})$; \tcp{Inverted Feedback}
     }
     %\For {$b_i$ from $\{ C_{j} \mid j \in [1, $\VariableStep$] \}$}{
     \For{$i\gets 0$ \textup{\textbf{ upto }} $k$ }{
     %\For {each bit $b_i$ from $0$ to $\VariableStep-1$}{
         $b'_i \leftarrow (\overline{m} \land b_i) \lor (m \land b_{n-k+i})$; \tcp{Forward Shift}
     }
    $O_{\text{next}}' \leftarrow O_{\text{next}} \lor (b_{n-1} \lor \overline{b_{n-1}'}) \land m $; \tcp{Overflow Checking}
 }
\end{algorithm}

\subsubsection{Input-Aware Rippling Minimization}
%Run-time Overflow Minimization
\label{sss:overflow_minimization}
Optimizing the \VariableStep-ary increments within a digit boosts performance, but the full carry propagation between digits remains a performance bottleneck. 
This section introduces Input-Aware Rippling Minimization (IARM), a run-time mechanism that postpones carry propagation to higher-order digits.
Recall, each digit is augmented with an additional flag bit, $O_{next}$.  Because this signals when there is a carry to propagate to the next digit, it increases the effective range of a digit from $2n-1$ to $4n-1$ unique values. Thus, when a value exceeds $2n-1$, we can avoid immediately triggering a carry.  Instead, the carry propagation can wait until a subsequent increment command could cause the counter to exceed $4n-1$. 

Based on this observation, we propose IARM, which selectively triggers \emph{carry rippling} based on increment history.  IARM is oblivious of the mask values stored in memory and must presume all count values are applied to at least one counter.  Thus, IARM implements a virtual counter which is incremented with all input values.  It uses this virtual counter to determine how long each counter can be delayed and issues carry commands just prior to increments that would make digits in the virtual counter exceed $4n-1$.  %

\begin{figure}[tbp]
    \centering
    \includegraphics[width=\linewidth]{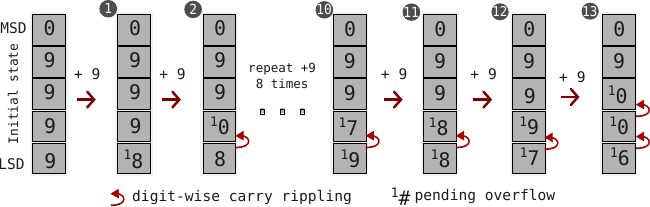}
    \vspace{-.2in}
    \caption{Increments with delayed overflow resolution.}
    %\vspace{-.2in}
    \label{fig:overflow_minimization}
    \vspace{-.2in}
\end{figure}

In Fig.~\ref{fig:overflow_minimization} we explain an example of how IARM avoids overflow ripple operations where the virtual counter has already been initialized to 9999.  Presuming a radix-10 ($n=5$) system with $\geq5$ digit counters, with traditional rippling, even an increment of 1 would cause a ripple effect from the lowest digit to the fifth digit.  The baseline approach must presume this type of ripple is possible in some counter at any time requiring ripple propagation in all digits for every increment.

In contrast, consider a series of increments by 9 in IARM. 
%In the bottom row, 
Digits represented as $^1$\# indicate digits storing values between $2n$ and $4n-1$ or between $^10$ and $^19$ for $n=5$. 
In step \solidcircle{1} no carry resolution is necessary because the lowest digit stores $^18$ and the entire counter is 999$^18$.  A second increment by 9 to step \solidcircle{2} does cause a ripple increment to the next digit, but then stops there resulting in 99$^10$7.  If we continue to add values of 9, we will eventually get to 99$^17$$^19$ in step \ding{191}, 99$^18$$^18$ in step \solidcircle{11} and 99$^19$$^17$ in step \solidcircle{12} prior to rippling beyond the tens digit in step \solidcircle{13} to 9$^10$$^10$$^16$.

Fig.~\ref{subfig:OVM} demonstrates the performance improvement from IARM over \VariableStep-ary increment only.
The \VariableStep-ary curves are reproduced from Fig.~\ref{subfig:k-ary} and require full carry propagation for each increment.  The levels for the state-of-the-art RCA method, which are determined by the worst-case maximum value, are also included.
As IARM is solely input-dependent (determined by the number of non-zero digits in one operand), varying counter capacity does not affect its operation count, as reflected by the single IARM curve, invariant of counter capacity.  IARM provides the fewest operation implementation over all other approaches particularly for radices 4--8.

\vspace{-.1in}
\subsection{Extending In-Memory Counting to NVMs}
\label{subsec:reram_counter}
This approach for building JCs in memory is extensible to other CIM approaches, as long as they provide a functionally complete set of bulk bitwise operations.
Considering \texttt{(N)AND} and \texttt{(N)OR} followed by writeback as primitive operations in nonstateful logic~\cite{pinatubo,yu2019enhanced}, the counting requires $3n+4$ operations, and the overflow checking additional $3$ operations, as illustrated by Fig.~\ref{subfig:pinatubo-listing}. 
However, MAGIC supports only \texttt{NOR}, which significantly increases the number of operations and, hence, the frequency of bit switching.
Fig.~\ref{subfig:magic-listing} shows the corresponding $\mu$Program to increment by 1 for MAGIC expressed in \texttt{NOR} operations. 
Specialized optimizations for MAGIC can be used to construct a program that requires only $6n+4$ operations for counting and overflow.  

In the next section we detail how these counting primitives, optimizations, and applicability to multiple CIM technologies can be used to realize the larger Count2Multiply architecture.

\begin{figure}[tbp]
    \centering
    \subcaptionbox{Pinatubo~\cite{pinatubo}\label{subfig:pinatubo-listing}}{\begin{minipage}{0.36\textwidth}\input{figures/ISCA/PinatuboListing}\end{minipage}    \vspace{-2ex}}%
    \hfill
    \subcaptionbox{MAGIC~\cite{magic}\label{subfig:magic-listing}}{\begin{minipage}{0.58\textwidth}\input{figures/ISCA/MagicListing}\end{minipage}    \vspace{-2ex}}%
    \vspace{-2ex}
    \caption{Pinatubo and MAGIC primitives for counting.}
\label{fig:Pinatubo_MAGIC}
\vspace{-0.2in}
\end{figure}

\section{Count2Multiply}
\label{count2multiply:presentation}

This section presents details of the Count2Multiply architecture for efficiently performing integer-binary matrix multiplication, addition, and other tensor-style operations.  
The architecture uses the host CPU (Fig.~\ref{fig:memory_array}) and, in particular, the Memory Controller (MCU) to issue memory command sequences to implement the in-memory high-radix counting from Sec.~\ref{sec:CIMCounters}.
The following sections detail this execution model, covering system-level integration (Sec.~\ref{subsec:exec-model}) and supported application kernels (Sec.~\ref{subsec:C2M-matmul}).  We continue to describe Count2Multiply with Ambit-style CIM, however, conceptually this approach can be extended to other CIM technologies (Sec.~\ref{subsec:reram_counter}).

\vspace{-.05in}
\subsection{Execution Model and System Integration}
\label{subsec:exec-model}

\begin{figure}[tbp]
\includegraphics[width=\columnwidth]{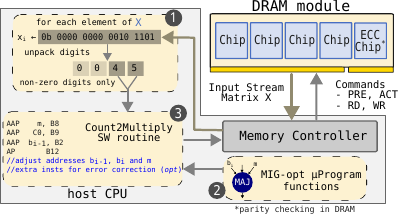}
\vspace{-0.2in}
\caption{CPU - DRAM system and $\mu$Program generation.}
\label{fig:memory_array}
\vspace{-0.2in}
\end{figure}

Count2Multiply employs a \emph{broadcast and accumulate} execution model, where a \emph{host-side program} converts an input stream into AAP/AP command sequences representing \emph{increments}, as illustrated in Fig.~\ref{fig:memory_array}. To generate these memory commands, i.e., %\texttt{AAP}/\texttt{AP} commands 
\ding{182}..\ding{184} in Fig.~\ref{fig:memory_array}.  \ding{182} The host-side program reads elements of $\mathbb{X}$ from the memory. 
 %It unpacks the input bits (stored in binary or any arbitrary encoding) into digits of the specified counter radix. 
After converting $X_i$ from binary to the counter radix, \ding{183} an increment sequence for each non-zero digit of $X_i$ based on the appropriate $\mu$Program (Fig.~\ref{subfig:microProgram}) is selected from an optimized CIM sequence template optimized by Majority synthesis~\cite{amaru2014majority}.  \ding{184} this $\mu$Program is issued by the MCU to one or many memory subarrays for parallel operation.  %The code for \emph{incrementing} and \emph{overflow checking} uses templates  (\ding{182}). 
An example of this process from Fig.~\ref{subfig:microProgram}: \ding{182} ``0b00101101'' is read from memory and unpacked into the digits ``45'' (radix-10). \ding{183} For each digit, we generate a \VariableStep-ary increment $\mu$Program with the row addresses (counter) of the corresponding digit position -- adding ``5'' to the ten's place and ``4'' to the hundred's place. 
Essentially, the input digit value determines the composition of \emph{forward shifts} and \emph{inverted feedback} (Sec.~\ref{sss:k-ary-inc}) as well as the row addresses for the corresponding counter-digit. %, which in turn increment the counters digit-by-digit accordingly.

\ding{184} These $\mu$Programs are then converted into a memory command sequence (i.e., \texttt{ACT/PRE} commands), which the MCU broadcasts to the memory chips%, as shown in \ding{184} in Fig.~\ref{fig:memory_array}
.
This enables selective incrementing of memory columns at destination addresses ($\mathbb{Y}$ values stored as \emph{counters} in memory) based on bit masks ($\mathbb{Z}$). In doing so, Count2Multiply updates only the necessary digits of $\mathbb{Y}$ to maintain computational accuracy, effectively implementing the \emph{early termination} of carry propagation among digits, consequently avoiding useless operations on high-order digits.

\noindent\textbf{IARM Integration.} The IARM mechanism (Sec.~\ref{sss:overflow_minimization}) is implemented on the host side.  Each $X_i$ value is first accumulated in a software-emulated \emph{virtual counter} before broadcasting the \VariableStep-ary \emph{increments} to memory.  This ensures correctness in the presence of potential overflows.  If accumulating the next $X_i$ would require a second overflow in one or more counters, overflow increment operations are issued as needed followed by the increment commands to accumulate $X_i$.  The host side implementation of IARM follows the steps detailed in Sec.~\ref{sss:overflow_minimization}.

\noindent\textbf{System Integration.} Fig.~\ref{fig:memory_array} illustrates a system where the CPU executes Count2Multiply’s routine, unpacking inputs and populating $\mu$Programs with designated rows to trigger counting directly in memory.
However, Count2Multiply is also compatible with non-CPU-based systems. For example, in FCDRAM~\cite{COTS-DRAM}, an FPGA can serve as the MCU, orchestrating $\mu$Program execution. Alternatively, a specialized control unit integrated into the MCU, like in SIMDRAM~\cite{SIMDRAM}, is also suitable for implementing the masked matrix accumulation program, thus independently populating $\mu$Programs for DRAM execution. 
The overhead of dynamically building $\mu$Programs is negligible, as the \texttt{AAP}/\texttt{AP} processing rate of the DRAM module is generally much lower than the $\mu$Programs generation on the host side, even when considering a single-core processor (see Sec.~\ref{sec:eval}).

\vspace{-.1in}
\subsection{Kernels Accelerated by Count2Multiply}
\label{subsec:C2M-matmul}
Count2Multiply offers a method for \emph{masked accumulation} in a wide, parallel manner, making it well-suited for tensor operations. 
%\noindent\textbf{Integer-Vector Binary-Matrix Operations.}
\subsubsection{Integer-Vector Binary-Matrix Operations}
\label{sec:vmmult}
Since counters increment a single input at a time, vector-matrix %and matrix-matrix 
multiplications is reinterpreted as \emph{masked matrix accumulations}.
In Fig.~\ref{fig:c2m-arch}a, we illustrate how vector-matrix multiplication $\vec{Y} = \vec{X}.\mathbb{Z}$ is formed by computing $\vec{Y} = \sum_{i=1}^{K} X_i.\vec{Z_i}$, wherein elements ($X_i$) of the integer-vector $\vec{X}$ are accumulated into the vector $\vec{Y}$, as predicated by the mask rows ($\vec{Z_i}$) in $\mathbb{Z}$.\footnote{In the general case, matrix $\mathbb{X}$ has shape $[M \times K]$, $\mathbb{Z}$ $[K \times N]$, and $\mathbb{Y}$ $[M \times N]$.} Here, $\vec{Y}$ is stored in memory as high-radix \emph{counters}, $\mathbb{Z}$ is stored in memory as binary masks, while $\vec{X}$ is an external input processed by the host CPU and issued by the MCU (Sec.~\ref{subsec:exec-model}). 
This approach draws inspiration from the sum of outer products to effectively implement an \emph{integer-binary} matrix multiplication through \emph{broadcast} and \emph{accumulation}. 

\subsubsection{Integer-Matrix Binary-Matrix Operations} Vector-matrix multiplication naturally generalizes to matrix-matrix multiplication when $M>1$\footnotemark[5] as shown in Fig.~\ref{fig:c2m-arch}a. Each row $\vec{Y}_o$ of the output matrix $\mathbb{Y}$ is computed independently as $\vec{Y}_o=\sum_{i=1}^{K} X_{oi}\cdot\vec{Z}_i,~\text{for } o = 1, \ldots, M$, where the matrix $\mathbb{Z}$
is reused. As the rows of $\mathbb{Y}$ are computed sequentially, each computed \textit{matrix} row can either be moved to a different subarray or used immediately. Copying the matrix row requires copying the memory rows dedicated to the high-radix counters to another subarray.  Recall these are memory rows to store the counters from the D-group in Fig.~\ref{fig:c2m-arch}b.  These memory rows can be reused for accumulation of the new matrix row for $\mathbb{Y}$.  This eliminates the need for dedicated counters for a specific row of $\mathbb{Y}$ within a single subarray and avoids the higher cost of copying the many more rows storing mask data, i.e., matrix $\mathbb{Z}$.

\subsubsection{Integer-Integer Matrix Operations} Integer-binary matrix multiplication can be extended to \emph{integer-integer} computation through \emph{bit-slicing} matrix $\mathbb{Z}$. To support low-precision $p$-bit \texttt{int} and \texttt{uint} matrices, each value is decomposed into canonical signed digit (CSD) form~\cite{canonical1,canonical2}, requiring $2 (p-1)$ or $p$ power-of-two-weighted binary masks, respectively. Each bit slice, representing a specific power-of-two significance and sign, maps to a row address in the memory subarray.  
The accumulation still uses counter array~$\mathbb{Y}$ but uses power-of-two-scaled inputs based on the bit slice value. We can think of each row of $\mathbb{Z}$ requiring multiple bitsliced memory rows.  The host-side routine scales the inputs based on the bit slice row address that indicates the power-of-two.  The host can also use shifting for scaling as the bitsliced masks represent only powers-of-two, avoiding the need for or use of a CPU multiplier to generate $\mu$Programs. 
The host side scaling allows accumulations for different bit slices to operate directly on a single counter row. 

\subsubsection{Additional Tensor Style Operations} Counting can also be leveraged to perform \emph{shift-left}, \emph{ReLU}, and \emph{vector addition} of counters. For $c << i$ the counter value can be added to itself $i$ times.  ReLU checks whether a counter is non-negative, which is possible by checking $O_{sign}$.  
Adding two vectors of counters ($C_1$ and $C_2$) is done with both counters in memory. 
This approach involves using one of the $n$-bit counters (e.g., $C_2$) as masks for unit incrementing a second counter ($C_1$) in place, i.e.,  $C_1 \gets C_1 + C_2 $. Algorithm~\ref{alg:johnson_counter_addition} illustrates the addition of pairs of counters by manipulating a counter $C_1$ based on the bits of another counter $C_2$. It first propagates the influence of high-order bits downwards using bitwise OR operations (Lines 2-4) and then refines the operation with a reverse pass using bitwise AND with the negated bits (Lines 6-8).

\begin{algorithm}[tbh]
\scriptsize
\caption{Johnson counter addition}
\label{alg:johnson_counter_addition}
\KwIn{ Counters $C_1$, $C_2$} 
$\Theta \gets C_2.MSB$\;
\For{$b$ in $[C_2.MSB, ..., C_2.LSB]$}{ 
    $mask \gets b \lor \Theta$\;
     increment$(1, C_1, mask)$; \tcp{Unit increment}
     }
  $\Theta \gets mask$\;
 \For{$b$ in $[C_2.LSB, ..., C_2.MSB]$}{ 
     $mask \gets \lnot b \land \Theta$\; 
     increment$(1, C_1, mask)$;\
 }
 \end{algorithm}

\section{Fault tolerance}
\label{sec:fault-tolerance}
High fault rates in CIM operations, especially in multi-row activation schemes, require a protection scheme (Sec.~\ref{subsec:reliab-bg}). 
For the multi-row activation primitive in DRAM, we 1) propose a general scheme that leverages existing ECC hardware, and 2) demonstrate it with our in-memory counters (Sec.~\ref{sec:CIMCounters}).
The core idea is to embed all CIM operations into ECC homomorphic operations.
In doing so, we exploit the fact that most faults in the intermediate steps will flip the final result, invalidating the ECC.
We show that cases where intermediate faults do not affect the final result -- making them undetectable via ECC -- are rare, occurring with very low probabilities due to CIM characteristics~\cite{yuksel2023pulsar,scoutinglogic}.

\begin{figure}[tbp]
\centering
\hfill
\subcaptionbox{\texttt{XOR} synthesis \label{subfig:XOR-Syn}}{\begin{tikzpicture}[
scale=0.8,
transform shape,
font=\small\ttfamily\bfseries,
roundnode/.style={circle, draw=MyOrange, fill=MyOrange!25, very thick, minimum size=0.8cm, inner sep=0},
innode/.style={rectangle, draw=MyDarkBlue, fill=MyDarkBlue!25, very thick, minimum size=.5cm},
outnode/.style={rectangle, draw=MyGreen, fill=MyGreen!25, very thick, minimum size=.4cm},
]
%Nodes

\node[roundnode]      (FR-AND)                           {AND};
\node[outnode] (IR1) [above left=0.6cm of FR-AND]{IR$_1$};
\node[outnode] (IR2) [above right=0.6cm of FR-AND]{IR$_2$};
\node[roundnode]      (IR1-OR)       [above = 0.2cm of IR1] {OR};
\node[roundnode]      (IR2-AND)       [above = 0.2cm of IR2]{AND};
\node[outnode, draw=MyDarkPurple, fill=MyDarkPurple!25] (FR) [below=0.2cm of FR-AND] {FR};
\node[innode] (A1) [above left=0.2cm of IR1-OR] {a};
\node[innode] (A2) [above left=0.2cm of IR2-AND] {a};
\node[innode] (B1) [above right=0.2cm of IR1-OR] {b};
\node[innode] (B2) [above right=0.2cm of IR2-AND] {b};
%Lines
\draw[->] (IR1-OR.south) -- (IR1.north);
\draw[->] (IR2-AND.south) -- (IR2.north);
\draw[->] (FR-AND) -- (FR);

\draw[->] (A1.south) -- (IR1-OR.north west);
\draw[->] (B1.south) -- (IR1-OR.north east);

\draw[->] (A2.south) -- (IR2-AND.north west);
\draw[->] (B2.south) -- (IR2-AND.north east);

\draw[{Circle[fill=white]}->] (IR2.south) -- (FR-AND.north east);
\draw[->] (IR1.south) -- (FR-AND.north west);
\end{tikzpicture}}%
\hfill
\subcaptionbox{Fault examples\label{subfig:XOR-TruthTable}}{\begin{tikzpicture}[
  font=\footnotesize\ttfamily,
  head color/.style args={#1/#2}{
    row 1 column #1/.append style={nodes={fill=#2}}},
  % swap order of row and column styles
  matrix/inner style order={
    every cell,
    even odd row, row,
    even odd column, column, 
    cell
  }
]

\matrix [
   matrix of nodes, nodes in empty cells,
   nodes={minimum width=0.5cm, align=center,
          minimum height=.3cm, text width = 0.5 cm, inner sep = 0, outer sep = 0,
          anchor=center,
          draw = black!5, thin},
   % add striped row style
   % every odd column/.style={nodes={fill=teal!15}},
   % every even column/.style={nodes={fill=teal!30}},
   % modify the feature column and header row
   column 1/.style={nodes={fill=MyDarkBlue!10}},
   column 2/.style={nodes={fill=MyDarkBlue!10}},
   column 3/.style={nodes={fill=MyGreen!10}},
   column 4/.style={nodes={fill=MyGreen!10}},
   column 5/.style={nodes={fill=MyDarkPurple!10}},
   row 1/.style={nodes={minimum height=0.5cm}},
   row 1 column 1/.style={nodes={fill=MyDarkBlue!75, draw=black}},
   row 1 column 2/.style={nodes={fill=MyDarkBlue!75, draw=black}},
   row 1 column 3/.style={nodes={fill=MyGreen!75, draw=black}},
   row 1 column 4/.style={nodes={fill=MyGreen!75, draw=black}},
   row 1 column 5/.style={nodes={fill=MyDarkPurple!75, draw=black}},
   row 1 column 6/.style={nodes={draw=black}},
   head color/.list={} % specify header colors
  ] (m)
  {
     \textbf{\texttt{a}} & \textbf{\texttt{b}}  & \textbf{\texttt{IR$_1$}} & \textbf{\texttt{IR$_2$}} & \textbf{\texttt{FR}} &
     \textbf{\texttt{XOR}}\\ 
     0 & 0 & 0 & 0 & 0 & 0\\
     0 & 1 & 1 & 0 & 1 & 1\\
     1 & 0 & 1 & 0 & 1 & 1\\
     1 & 1 & 1 & 1 & 0 & 0\\
     0 & 0 & \textcolor{red}{1$^*$} & 0 & 1 & 0\\%1fault detect
     1 & 1 & \textcolor{red}{0$^*$} & 0 & 0 & 0\\%1 fault undetect
     0 & 1 & \textcolor{red}{0$^*$} & \textcolor{red}{1$^*$} & 0 & 1\\% two fault detect
     0 & 1 & \textcolor{red}{0$^*$} & \textcolor{red}{1$^*$} & \textcolor{red}{1$^*$} & 1\\%three fault undetect
    };
\node[fit={(m-1-1.north west) (m-1-6.south east)}, inner sep = 0, draw=black, thick]{};
\node[fit={(m-1-1.north west) (m-9-2.south east)}, inner sep = 0, draw=black]{};
\node[fit={(m-1-3.north west) (m-9-4.south east)}, inner sep = 0, draw=black]{};
\node[fit={(m-1-5.north west) (m-9-6.south east)}, inner sep = 0, draw=black]{};

%\node[fit={(m-2-4.north west) (m-2-4.south east)}, inner sep = 0, draw=black, very  thick, rounded corners=.15cm]{};
%\node[fit={(m-5-3.north west) (m-5-3.south east)}, inner sep = 0, draw=black, very thick, rounded corners=.15cm]{};

\end{tikzpicture}}%
\hfill
\vspace{-0.15in}
\caption{Synthesis of \texttt{XOR} function via CIM gates.}
\label{fig:xor-cim}
\vspace{-.1in}
\end{figure}

\vspace{-.1in}
\subsection{Fault Protection Scheme Setup}
\label{subsec:fault-setup}
Existing ECC schemes for memory are not homomorphic over \texttt{AND} or \texttt{OR} operations~\cite{ambit,CORUSCANT}.
However, many commercially used ECCs, including Hamming, Reed-Solomon, and BCH codes, are homomorphic over \texttt{XOR}. 
We embed our CIM operations in \texttt{XOR} to enable the use of traditional ECCs once \texttt{XOR} is calculated. 
The synthesis of an \texttt{XOR} function using CIM primitives is a two-step process, as shown in Fig~\ref{subfig:XOR-Syn}. The operation to protect, be it \texttt{OR} or \texttt{AND}, generates, \texttt{IR$_1$} or \texttt{IR$_2$}, respectively.  From the error correction perspective this an intermediate result (IR).  However, we also compute the complementary IR and use \texttt{IR$_1$} and \texttt{IR$_2$} that generate the final \texttt{XOR} result, \texttt{FR}.  FR can detect errors that could have occurred when computing \texttt{IR$_1$},  \texttt{IR$_2$}, and/or \texttt{FR} itself.

Two properties from this set of instructions exist.  Most faults from an intermediate gate will flip \texttt{FR}, invalidating the parity bits for \texttt{FR}.  The fault modes that would not be detectable are from a data-dependent pattern wherein the success rate is better than traditional DRAM accesses~\cite{yuksel2023pulsar, scoutinglogic}. 
In \texttt{MAJ3} gates~\cite{ambit}, these cases result in charge-sharing across three rows of `1's or `0's such that sensing margins are 
equal to or greater than those of a standard read.
In addition, the \texttt{NOT} operation is equivalent to \texttt{XOR} with `1', which can be directly protected by ECC in a manner functionally equivalent to reading~\cite{ambit,pinatubo}.
Thus, we ensure fault protection to guard against all \textit{likely} CIM faults.

We observe that there are three possible fault combinations:
\ding{172} a single fault could occur in any result (\texttt{IR$_1$},\texttt{IR$_2$}, or \texttt{FR}); \ding{173} two faults could occur: with both in \texttt{IR$_1$} and \texttt{IR$_2$} or with one in an IR and one in \texttt{FR}; 
\ding{174} a fault occurs in all three CIM calculations.
We illustrate the impact of each case with an example in Fig.~\ref{subfig:XOR-TruthTable}.
Recall, our proposed scheme will protect cases that flip the \texttt{FR} and thus invalidate the parity check.  Fig.~\ref{subfig:XOR-TruthTable} shows the single bit equivalent of comparing \texttt{FR} to the actual \texttt{XOR} result with faults in \textcolor{red}{red$^*$}. 
An example of case \ding{172} is shown in the fifth row of Fig.~\ref{subfig:XOR-TruthTable}.  A fault occurs flipping \texttt{IR$_1$} to `1'.
The fault propagates to \texttt{FR} and is detected by parity check (i.e., \texttt{FR} $\neq$ \texttt{XOR}).
If a single \textit{unlikely} fault occurs, our scheme cannot detect it as illustrated in the sixth row (fault in \texttt{IR}$_1$ but \texttt{FR} $=$ \texttt{XOR}) making it a silent error.
In case \ding{173}, shown in the seventh row of Fig.~\ref{subfig:XOR-TruthTable} a fault occurs in both \texttt{IR$_1$} and \texttt{IR$_2$}. 
Because neither of these faults were \textit{unlikely}, the fault is propagated to \texttt{FR} and $\neq$ \texttt{XOR}, making it detected.
In case \ding{174}, the scheme will not protect against faults in all three operations, shown in the eighth row of Fig.~\ref{subfig:XOR-TruthTable}.
The only \textit{likely} undetectable faults occur with one or both \texttt{IR}s being faulty \textit{and} faulty computing of \texttt{FR}.  Recomputing \texttt{FR} correctly reveals the~fault.

\vspace{-.1in}
\subsection{Fault Tolerant In-Memory Counting}
\label{subsec:integrating-fault-tolerance}
\begin{figure}[tbp]
\centering
\subcaptionbox{Generic protected \emph{forward shift} $\mu$Program\label{subfig:protect-listing}}{\input{figures/FaultTolerance/ProtectionListingGeneric}}%
\hfill
\subcaptionbox{Fault propagation and detection in masking operation\label{subfig:protect-example}}{\begin{tikzpicture}[
  font=\fontsize{7}{7}\selectfont\ttfamily,
  head color/.style args={#1/#2}{
    row 1 column #1/.append style={nodes={fill=#2}}},
  % swap order of row and column styles
  matrix/inner style order={
    every cell,
    even odd row, row,
    even odd column, column, 
    cell
  }
]

\matrix [
   matrix of nodes, nodes in empty cells,
   nodes={ align=center, minimum width = .40cm, text width=0.40cm, minimum height=.50cm, anchor=center, inner sep  =0,outer sep = 0,draw=black!5},
   % add striped row style
   % every odd column/.style={nodes={fill=teal!15}},
   % every even column/.style={nodes={fill=teal!75}},
   % modify the feature column and header row
   column 1/.style= {nodes={ minimum width = 0.50cm, text width=0.50cm, draw=black}}, %inner ysep =0
   row 2 column 1/.style={nodes={fill=MyDarkBlue!75}},
   row 3 column 1/.style={nodes={fill=MyDarkBlue!75}},
   row 4 column 1/.style={nodes={fill=MyGreen!75}},
   row 5 column 1/.style={nodes={fill=MyGreen!75}},
   row 6 column 1/.style={nodes={fill=MyDarkPurple!75}},
   row 2/.style={nodes={fill=MyDarkBlue!25}},
   row 3/.style={nodes={fill=MyDarkBlue!25}},
   row 4/.style={nodes={fill=MyGreen!25}},
   row 5/.style={nodes={fill=MyGreen!25}},
   row 6/.style={nodes={fill=MyDarkPurple!25}},
   row 5 column 4/.style={nodes={fill=MyRed!50}},
   column 10/.style={nodes={fill=MyPink!15}},
   row 4 column 10/.style={nodes={pattern=north west lines, pattern color=MyPink!75}},
   row 5 column 10/.style={nodes={pattern=north west lines, pattern color=MyPink!75}},
   row 6 column 10/.style={nodes={fill=MyRed!50}},
   row 7 column 4/.style={nodes={fill=MyRed!50}},
   row 1/.style= {nodes={ text=black, draw=black}},
   row 1 column 1/.style={nodes={fill=none, draw=none}},
   head color/.list={10/MyPink!75} % specify header colors
  ] (n)
  {
                            &   &   &   &   &   &   &   &   & \textbf{P}\\
  \textbf{\texttt{m}}                   & 1 & 1 & 1 & 1 & 0 & 0 & 0 & 0 & 0\\
  \textbf{\texttt{b$_\texttt{j}$}}      & 0 & 1 & 1 & 1 & 0 & 1 & 0 & 1 & 0\\
  \textbf{\texttt{IR$_1$}}              & 1 & 1 & 1 & 1 & 0 & 1 & 0 & 1 & 0\\
  \textbf{\texttt{IR$_2$}}              & 0 & 1 & 0 & 1 & 0 & 0 & 0 & 0 & 0\\
  \textbf{\texttt{FR}}                  & 1 & 0 & 1 & 1 & 0 & 1 & 0 & 1 & 0\\
  };

%add multicolumn header by "fit" library
\node[fit={(n-1-2.north west) (n-1-9.south east)},fill=MyGray, inner sep = 0, text depth=0.2ex, minimum height=.50cm, text=black, draw=black]{\textbf{Data Row}};
% Add emphasis on selection by the use of "fit" library
\node[fit={(n-1-2.north west) (n-1-10.south east)},
       thick, inner sep=0pt,
      draw=black]{};
\node[fit={(n-2-1.north west) (n-6-1.south east)},
     thick, inner sep=0pt,
      draw=black]{};
\node[fit={(n-2-2.north west) (n-6-10.south east)},
     thick, inner sep=0pt,
      draw=black]{};
\node[fit={(n-5-4.north west) (n-5-4.south east)},
      ultra thick, inner sep=0pt, rounded corners=1mm,
      draw=red]{};
\node[fit={(n-6-4.north west) (n-6-4.south east)},
      ultra thick, inner sep=0pt, rounded corners=1mm,
      draw=red]{};
\node[fit={(n-6-10.north west) (n-6-10.south east)},
      ultra thick, inner sep=0pt, rounded corners=1mm,
      draw=red]{};

\node[right =.05cm of n-5-4.north east, inner sep = 0pt, outer sep = 0pt, text = red] {\textbf{1}};
\node[right =.05cm of n-6-4.north east, inner sep = 0pt, outer sep = 0pt, text = red] {\textbf{2}};
\node[right =.05cm of n-6-10.north east, inner sep = 0pt, outer sep = 0pt, text = red] {\textbf{3}};
% \node[fit={(n-7-4.north west) (n-7-4.south east)},
%       ultra thick, inner sep=0pt, rounded corners=1mm,
%       draw=red]{};
%\draw[->] (n-5-4.south) --  (n-7-4.north);
%\draw[->] (n-7-4.east) --  (n-7-10.west);

\end{tikzpicture}}%
\vspace{-0.15in}
\caption{ECC scheme for in-memory counting.}
\label{fig:ProtectEx}
\vspace{-.2in}
\end{figure}

In our counting method, we compute each new row by combining two rows (positions $i$ and $i-k$) masked with $m$ and $\overline{m}$. Here, an \texttt{OR} operation is functionally equivalent to \texttt{XOR}, presuming an \texttt{XOR} can be directly computed on the parity bits~\cite{zhou2022flexidram}. This is because masked locations are set to `0', so `1's can only be stored in unmasked locations, ensuring that `1's are mutually exclusive between the two rows. This method is useful for Count2Multiply because it only works on mutually exclusive masked rows that are created for \VariableStep-ary counting.  Thus, only \texttt{AND} for masking must be protected by embedding the masking in computing \texttt{XOR} as described in Sec.~\ref{subsec:fault-setup}.

Fig.~\ref{subfig:protect-listing} details the generic increment $\mu$Program when using our protection scheme.
In line 2 we compute one of the masking results we want to protect: $b_j \land m$. %\texttt{b$_\texttt{j}$ $\land$ m}.
In lines 3 and 4 we perform the additional operations required to complete an \texttt{XOR} for protecting the result.
In line 5 we copy the masking result to a temporary destination so that \texttt{IR$_1$} and \texttt{IR$_2$} can be reused in this example.
In lines 6--8 we compute and protect the masking operation $b_i \land \overline{m}$ %\texttt{b$_\texttt{i} \land \overline{\texttt{m}}$}.
Finally we update the row with our protected results (line 11).
The listing demonstrates the operation overhead to enable traditional ECC checks to detect errors in CIM results.

%\textcolor{blue}{
Fig.~\ref{subfig:protect-example} shows fault protection for an example row during in-memory masking.
Initially, mask $m$ %\texttt{m} 
and JC bit $b_j$ %\texttt{b}$_\texttt{j}$ 
rows are stored with a traditional parity bit, \texttt{P}.
A fault occurs during the masking step which generates \texttt{IR$_2$}.
This fault flips the bit stored in the position highlighted and labeled \textcolor{red}{\textbf{\texttt{1}}}.
For the error detection scheme we compute \texttt{IR$_1$} and \texttt{FR}, noting the flip from \textcolor{red}{\textbf{\texttt{1}}} propagates to \textcolor{red}{\textbf{\texttt{2}}}.
\texttt{FR} fails the parity check \textcolor{red}{\textbf{\texttt{3}}} performed in ECC hardware. 
%In this case, the parity check will fail (\textcolor{red}{\textbf{\texttt{3}}}), signaling a fault occurred that 
This requires repeating the computation. 
In Fig.~\ref{subfig:protect-listing}, this implies that if an ECC check fails after line 4 it is necessary to restart from line~2.

\begin{table}[tbh]
    \centering
    \setlength\tabcolsep{1.5pt}
    \scriptsize
    \vspace{-.1in}
        \caption{Effects of varying number of \texttt{FR} checks with different inherent CIM fault rates.}
        \vspace{-.05in}
    \begin{tabular}{c|lll|lll|lll}
        \toprule
        \texttt{FR} checks & \multicolumn{3}{c|}{2} & \multicolumn{3}{c|}{4} & \multicolumn{3}{c}{6} \\
        Fault rate & $10^{-1}$ & $10^{-2}$ & $10^{-4}$ &$10^{-1}$ & $10^{-2}$ & $10^{-4}$ &$10^{-1}$ & $10^{-2}$ & $10^{-4}$ \\
        \midrule
        Error rate &1.4E-3 & 1.5E-6 & 1.5E-12 & 1.4E-5 & 1.5E-10 & \emph{1.0E-20} & 1.4E-7 & 1.5E-14 & \emph{1.0E-20} \\
        Detect rate &3.1E-1 &3.5E-2 & 3.5E-4 & 4.4E-1 & 5.4E-2 & 5.5E-4 & 5.5E-1 & 7.3E-2 & 7.5E-4\\
        Ambit & \multicolumn{3}{c|}{$13n+16$} & \multicolumn{3}{c|}{$23n + 26$} & \multicolumn{3}{c}{$33n + 36$}\\
        \bottomrule
    \end{tabular}
    \label{tab:fault} 
    \vspace{-.1in}
\end{table}

\subsection{Optimizations and Extensions}
While performing the increment step with inversion $\overline{b_i}\land m$, the protection can be combined with that of $b_i\land \overline{m}$ by using De Morgan's Laws (they produce valid \texttt{IR$_1$}, \texttt{IR$_2$} for \texttt{XOR} synthesis as in Fig.~\ref{subfig:XOR-Syn}). 
%\textcolor{blue}{
Because we can protect two masking steps through computing \texttt{XOR} when performing inverted feedback and the inverted feedback step accounts for half of all increment steps (Fig.~\ref{fig:k-ary}), the net protection overhead can be reduced by 25\%. %}
Further, in Ambit~\cite{ambit} the special address groupings can be used to perform the \texttt{NOT} simultaneously with other operations.
However, Ambit suffers seriously from the requirement that CIM operations be performed within the limited set of CIM-enabled addresses.
Additionally, DRAM CIM solutions in general~\cite{ambit, COTS-DRAM} are destructive, requiring operand copying before each operation.
For all schemes, we carefully consider potential optimizations and challenges in order to make a fair comparison.
The operation count for each scheme with different levels of protection is given in Tab.~\ref{tab:fault}.
While the single error detection scheme can detect all possible single CIM faults, this may not suffice for higher fault rates, i.e., $10^{-1}$ -- $10^{-4}$. Our protection scheme can be extended to two error detection and beyond.
%\textcolor{blue}{
Additionally, the number of times we repeat the \texttt{FR} is configurable.
Since all cases of \textit{likely} faults going undetected require one of the faults to be in the \texttt{FR} (Sec.~\ref{subsec:fault-setup}), we can simply repeat the \texttt{FR} computation to improve our fault tolerance.
Tab.~\ref{tab:fault} compares protection schemes with varying repeats (\texttt{FR checks}) applied to different CIM fault rates.
The \textit{error} and \textit{detect} rates report the per-bit probability of undetectable and detectable errors respectively. 
In our analysis, we assume that the \emph{unlikely} CIM faults will exhibit a fault rate similar to that of a typical memory read operation (conservatively estimated at $10^{-20}$ for DRAM~\cite{DRAMerrorsinwild}), giving an upper bound on our fault tolerance level.
The error rates bounded by DRAM fault rates in Tab.~\ref{tab:fault} are italicized.

\begin{table}[h]
\vspace{-.1in}
\centering
\caption{Memory organization and architectural parameters.}
%\vspace{-.05in}
\vspace{-0.15in}
\centering
\scriptsize
\label{tab:Parameter}
\begin{tabular}{c|c}
\toprule
%Technology & Parameter & Value \\
%\midrule
\multirow{3}{*}{DRAM}& DDR5\_4400, 1 channel, 1 rank, 8 devices + ECC;\\
& 4Gb DRAM chip, 32 banks, 1 kB row size, 1024 rows per subarray \\
\midrule
\multirow{2}{*}{host CPU} &  1 out-of-order core @ 3 GHz (instruction generation and offloading);	 \\
& Memory Controller 8 kB row size, FR-FCFS scheduling \\  
%\midrule
%\multirow{2}{*}{RTM} &  \SI{8}{\giga\byte}, 64 banks, 512 subarrays per bank	 \\
% & \# subarrays    			& 64 x 512 = 32768 \\ 
%&   Rows and columns per subarray:  512, 4096 \\
\bottomrule
\end{tabular}
\vspace{-.15in}
\end{table}

% \vspace{-0.1in}
\section{Experimental Setup and Results} 
\label{sec:eval}
To quantitatively evaluate the Count2Multiply approach, Ambit-style DRAM CIM is simulated by extending NVMain/RTSim~\cite{nvmain2.0,rtsim} with a cycle-level CIM simulation model. Our implementation of Ambit and SIMDRAM was rigorously validated against the results reported in~\cite{ambit,SIMDRAM} and by MIMDRAM's simulator~\cite{oliveira2024mimdram}. 
%The DRAM-based Count2Multiply is built on Ambit\jp{unnecessary}.
%The DRAM- and RTM-based Count2Multiply are built on Ambit and CORUSCANT~\cite{CORUSCANT}, respectively.
% We compare our design to a high-end GPU and a state-of-the-art in-DRAM design~\cite{SIMDRAM}.
The architectural parameters of our memory organization are listed in Tab.~\ref{tab:Parameter}. 
%The energy and latency parameters are the same as in the baseline CIM systems~\cite{SIMDRAM, CORUSCANT}. 
Our setup follows commercial DRAM organization and timing.% constraints. %, including critical parameters affecting parallelism. %, e.g., $t_{FAW}$ and $t_{RRD}$. %Subarray-level parallelism is not considered. %We are also not considering any subarray level parallelism. 

\begin{figure*}[h]
    \centering
    \subfloat {
    \includegraphics[width=0.65\columnwidth]{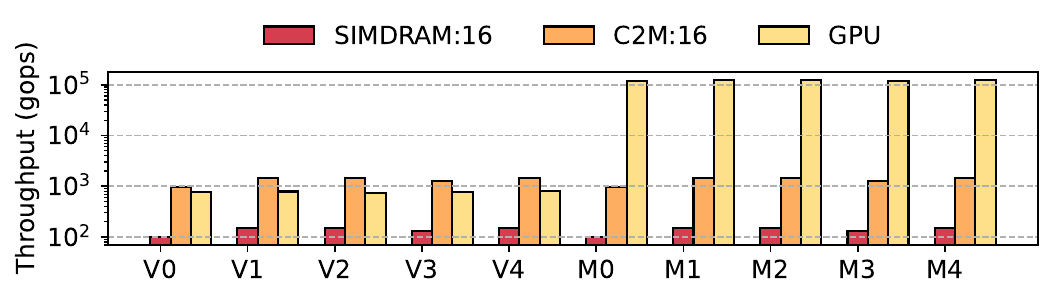}
    \label{subfig:gpu:perf}
    }
    \hfill
    \subfloat {
    \includegraphics[width=0.67\columnwidth]{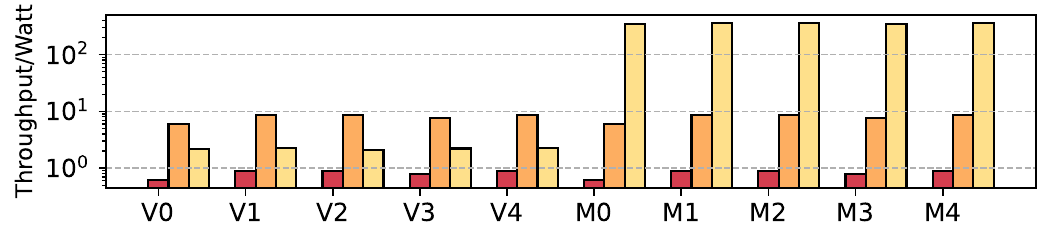}
    \label{subfig:gpu:perf}
    }
    \hfill
    \subfloat{
    \includegraphics[width=0.67\columnwidth]{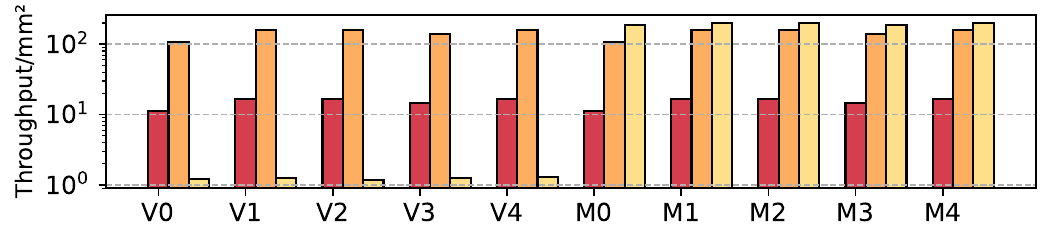}
    \label{subfig:gpu:perf}
    }
    \vspace{-0.15in}
    \caption{Performance comparison for real ternary GEMM and GEMV~\cite{touvron2023llama,touvron2023llama2,ma2024era}.}
    \label{fig:gpu}
\vspace{-0.15in}
\end{figure*}

\subsection{Configurations and Workloads}
\label{subsec:config}
We compare Count2Multiply %across technologies 
with state-of-the-art in-memory adders and a GPU as follows:
\begin{itemize}
\item \emph{SIMDRAM:X} RCA-based CIM design~\cite{SIMDRAM} using X banks. 
\item \emph{GPU:} NVIDIA RTX 3090 Ti -- 328 Tensor Cores (TCs). Each data point averages ten runs with a warm-up phase to avoid cold cache effects. GPU kernel performance and power reported with cudaEvents API and nvidia-smi, excluding data transfer. GPU area is 628 $mm^{2}$~\cite{nvidia3090whitepaper}.
\item \emph{C2M:X} is DRAM-based Count2Multiply using X banks.

\end{itemize}

\label{subsec:benchmarks}

\noindent We evaluate Count2Multiply on the following workloads:

\noindent \textit{GEMV and GEMM}:
We use the GEMM and GEMV shapes (M, N, and K) from Tab.~\ref{table:gemm_gemv} derived from~\cite{touvron2023llama, touvron2023llama2}. These shapes represent the key computational loads in the models and serve as effective proxies for assessing their performance. 

\begin{table}[tbp]
\centering
\scriptsize
\caption{GEM\textbf{V} and GEM\textbf{M} dimensions from~\cite{touvron2023llama, touvron2023llama2}.}
\vspace{-.08in}
\begin{tabular}{c | c c c c |c c c c }
\toprule
Model & ID & M & N & K & ID  & M & N & K \\ 
\midrule
LLaMA & V0 & 1 & 22016 & 8192 & M0 & 8192 & 22016 & 8192 \\ 
LLaMA &  V1 & 1 & 8192 & 22016 & M1 & 8192 & 8192 & 22016 \\ 
LLaMA-2 & V2 & 1 & 8192 & 8192 & M2 & 8192 & 8192 & 8192 \\ 
LLaMA-2 & V3 & 1 & 28672 & 8192 & M3 & 8192 & 28672 & 8192 \\ 
LLaMA-2 & V4 & 1 &  8192 & 28672 & M4 & 8192 & 8192 & 28672 \\ 
\bottomrule
\end{tabular}
\label{table:gemm_gemv}
% \vspace{-.15in}
\end{table}

\noindent\textit{Transformer networks: }
The attention mechanism faces memory-bound challenges due to its large-scale data requirements. We evaluate all GEMM operations in the attention layer of the BERT model. % as they are the most time-consuming component. 
Ternary parameters are also considered~\cite{chee2024quip}.

\noindent\textit{Pre-alignment filtering}
is a memory-intensive step in DNA analysis, where the input genome is compared to a reference genome stored as bitvectors in memory~\cite{GRIM_2018}. Nucleotide repetitions in the \emph{reads} of input genomes are represented as integers. For our evaluation, we use a human genome and a similar setup to prior work~\cite{ALPHA_2021}.

\noindent\textit{TWNs: }
Convolutional layers are computationally most intensive in modern NNs. 
In our experiments, we consider LeNet, VGG-13 and VGG-16 models to facilitate comparison with prior works. 

\noindent\textit{Graph Convolutional Networks} (GCNs) operate on graph-structured data by aggregating and transforming features from a node’s local neighborhood using the graph's connectivity. We evaluate GCNs on a node classification task using the PubMed dataset~\cite{chen2023bitgnn}.

\begin{figure}[tbp]
    \centering
%    \vspace{-0.15in}
    \subfloat[Latency] {
    \includegraphics[height=1.05in]{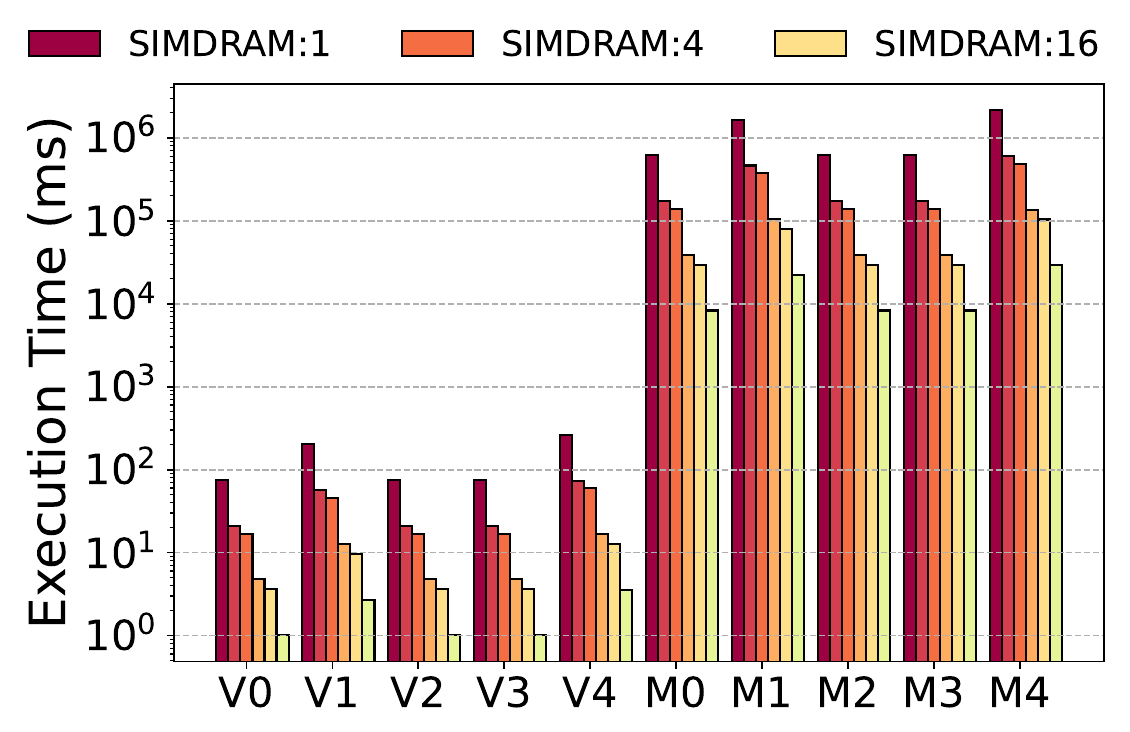}
    \label{subfig:gemm_gemv:exectime}
    }
    \subfloat[Throughput] {
    \includegraphics[height=1.05in]{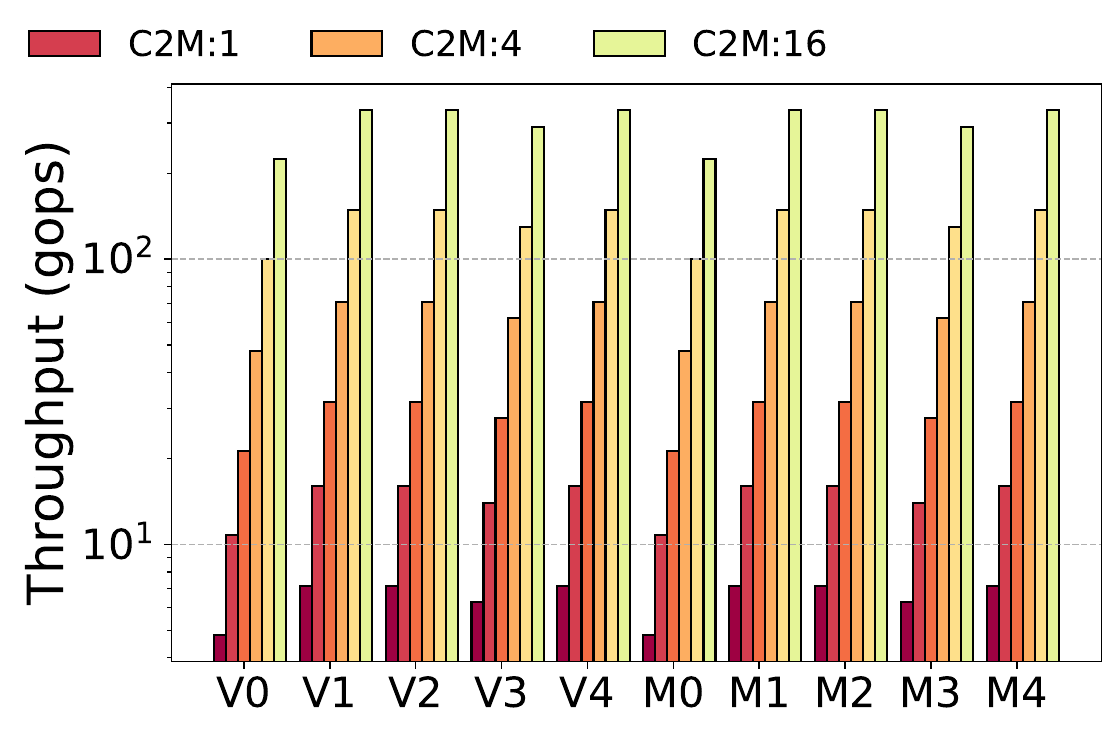}
    \label{subfig:gemm_gemv:exectime}
    }
    \vspace{-0.15in}
    \caption{Comparison of DRAM designs on ternary GEMV/GEMM from LLAMA and LLAMA-2 models~\cite{touvron2023llama,touvron2023llama2,ma2024era}.}
    \label{fig:gemm_gemv:all}
    \vspace{-0.15in}
\end{figure}

\vspace{-0.08in}
\subsection{Design Space Exploration: LLM Kernels}
%\subsection{Comparison to DRAM-Based CIM and GPU}
\label{subsec:comp-simdram}

In this section we evaluate Count2Multiply on tensor kernels from large language models LLaMA and LLaMA-2 for different metrics.
%conduct a full system\ak{full system? Should we instead say: we conduct evaluation across deferent metrics?} evaluation of Count2Multiply.

\subsubsection{Impact of DRAM Parallel Execution} %\jp{Use noindent and bold text instead of subsubsection will reduce some space}
Fig.~\ref{fig:gemm_gemv:all} presents a comparative analysis of the performance, i.e., latency, throughput and throughput per Watt, of only in-DRAM implementations with different subarray CIM parallelism for integer-ternary GEMV and GEMM workloads described in Tab.~\ref{table:gemm_gemv}. The evaluation uses an 8-bit signed integer input and radix-4 counters.  
All configurations assume an accumulation capacity of 64-bit integers to ensure computational precision. 

Due to the sequential nature of RCA in \emph{SIMDRAM}, \emph{C2M} consistently outperforms \emph{SIMDRAM} on all workloads and all system configurations. On average (geomean), \emph{C2M} is $2\times$ faster and delivers $1.15\times$ higher throughput and throughput per Watt. 
This finding is in agreement with previous results obtained for a single addition (Fig.~\ref{fig:performance}) and confirms that \emph{C2M} maintains its performance gains in more complex kernels such as GEMV and GEMM. 

We vary the number of banks from one to a maximum of 16. With a single bank, each step's latency is considerably high, dictated by the $t_{AAP}$ and $t_{RRD}$ timings, allowing one \texttt{AAP} operation every $t_{AAP}$ + $t_{RRD}$. With 4 banks, we can overlap four \texttt{AAP} commands across different banks, each separated by $t_{RRD}$. However, the delay between the first and fifth activation commands is still limited by $t_{AAP}$ + $t_{RRD}$ since the fifth can only start after the first finishes. For 16 banks, although constrained by the four activation window (FAW), the latency between the first and fifth activation is now bounded by $t_{tFAW}$, which is shorter than $t_{AAP}$ ($t_{RAS}$ + $t_{RP}$ + 4).

\begin{comment}[tbp]
    \centering
    \subfloat[Throughput] {
    \includegraphics[width=0.3\textwidth]{figures/results/GPU/gpu_normalized_throughput.pdf}
    \label{subfig:gpu:perf}
    }
    \subfloat[Performance per Watt] {
    \includegraphics[width=0.3\textwidth]{figures/results/GPU/gpu_normalized_energy.pdf}
    \label{subfig:gpu:perf}
    }
    \subfloat[Performance per Area] {
    \includegraphics[width=0.3\textwidth]{figures/results/GPU/gpu_normalized_performance_area.pdf}
    \label{subfig:gpu:perf}
    }
    % \vspace{-0.15in}
    \caption{GPU-normalized performance for real ternary GEMM and GEMV~\cite{touvron2023llama,touvron2023llama2,ma2024era}.}
    \label{fig:gpu}

\end{comment}

\subsubsection{Comparison with GPU}
\label{subsec:c2m-comp-gpu}
Fig.~\ref{fig:gpu} presents throughput and throughput per Watt and area of \emph{SIMDRAM} and \emph{C2M}, all normalized to the GPU baseline. 
As expected, with GPUs and BLAS routines being particularly designed and hand-optimized for GEMM, the CIM accelerators exhibit lower throughput than the GPU. 
Note that all results for in-DRAM designs use a single rank with one subarray per bank doing the computations. The results scale linearly with increasing the number of CIM subarrays and ranks. Further, we are using conservative estimates with a $t_{FAW}$ of \SI{14.5}{\nano\s}. All-bank activation, as suggested in prior work in the CIM domain~\cite{paik2022achieving}, will clearly lead to superior throughput compared to GPU in all configurations, it incurs higher power consumption. 

\begin{figure}[h]
%\vspace{-.05in}
\centering
    \includegraphics[width=\columnwidth]{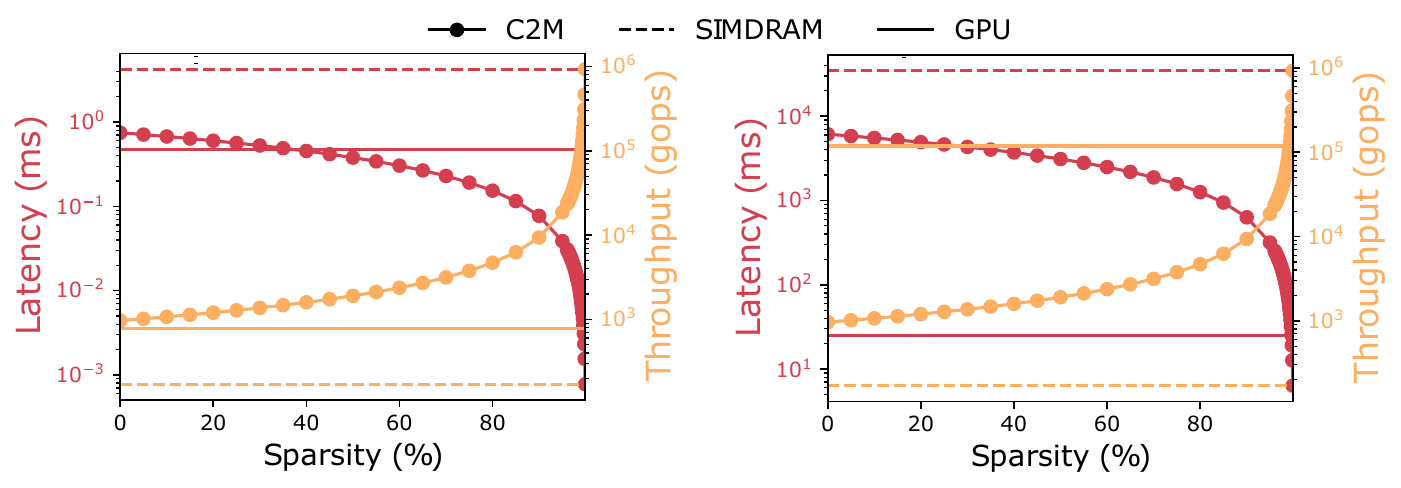}
    \vspace{-0.18in}
    \caption{Performance using sparse inputs: (left) Vector-Matrix Multiply (V0), (right) Matrix-Matrix Multiply (M0). }
    \label{fig:sparse}
    
\end{figure}

\subsubsection{Impact of Sparsity on Performance}

As described in Sec.~\ref{subsec:exec-model}, Count2Multiply skips zero-value inputs (and also zero digits from non-zero-value inputs), making it an ideal fit for sparse matrix operations which are common across various domains, including graph-based workloads, scientific computing, and deep learning~\cite{gao2023systematic}. In deep learning, pruning techniques can reduce the number of parameters in neural networks by up to 99\%~\cite{hoefler2021sparsity}. However, in other applications, e.g., graph neural networks, the inherent sparsity ranges from 90\% to 99.5\%~\cite{qiu2021optimizing}. 

\begin{figure}[tbp]
    \vspace{-0.15in}
%\vspace{-.05in}
\centering
    \subfloat[DNA Filtering] {
    \includegraphics[width=0.48\columnwidth]{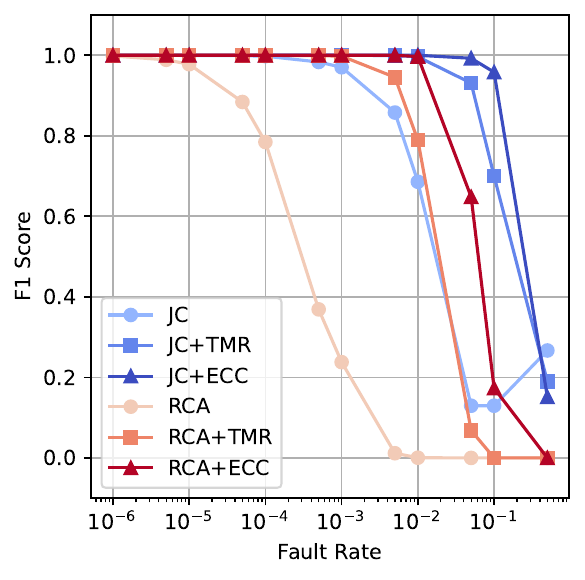}
    \label{subfig:filter:falt}
    }
    \subfloat[BERT] {
    \includegraphics[width=0.48\columnwidth]{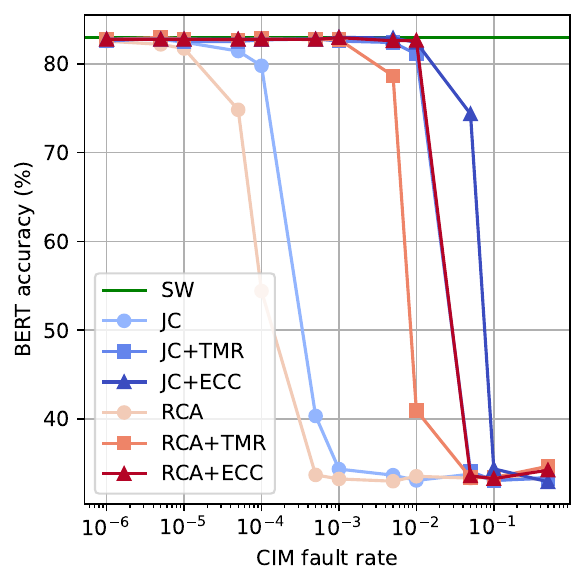}
    \label{subfig:bert:fault}
    }
    \vspace{-0.15in}
    \caption{Accuracy comparison under CIM fault probability.}
    \label{fig:fault}
\end{figure}

\begin{figure*}[tbp]
    \centering
    \subfloat {
    \includegraphics[width=0.66\columnwidth]{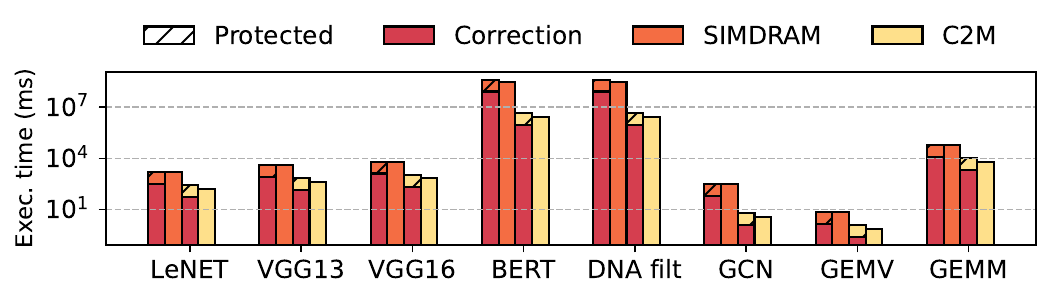}
     \label{subfig:use_cases:lat}
    }
    \hfill
    \subfloat {
    \includegraphics[width=0.66\columnwidth]{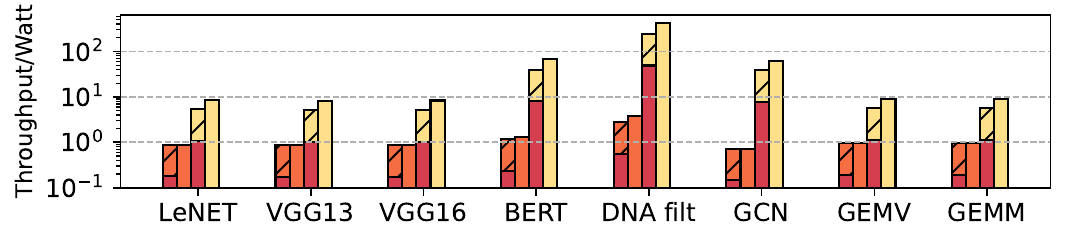}
    \label{subfig:use_cases:power}
    }
    \hfill
    \subfloat{
    \includegraphics[width=0.66\columnwidth]{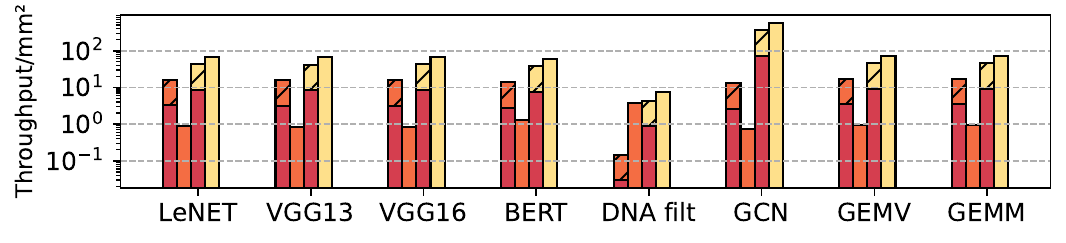}
    \label{subfig:use_cases:area}
    }
    \vspace{-0.15in}
    \caption{Performance comparison of real-world workloads, including the protection scheme overhead. 
    }
    \label{fig:use_cases}
    % \vspace{-0.15in}
\end{figure*}

Fig.~\ref{fig:sparse} compares the latency (including memory transfer) and throughput of GPU, SIMDRAM, and DRAM-based Count2Multiply (C2M) with a 16-bank configuration. We vary the degree of sparsity from 0\% to 99.9\% in V0 and M0 workloads (Tab.~\ref{table:gemm_gemv}), with latency (ms) shown on the primary vertical axis and throughput (GOPS) on the secondary axis. Using the GPU as the baseline, C2M outperforms SIMDRAM by orders of magnitude. Compared to the GPU, it delivers comparable latency and even surpasses it beyond $\sim$40\% sparsity in GEMV and 99.6\% in GEMM. Additionally, C2M outperforms GPU throughput in GEMV even for full dense inputs and exceeds GPU throughput at 99.1\% sparsity in the GEMM workload. Note that the GPU baseline benefits from optimized tensor cores and cuBLAS, while the C2M gains are mainly due to our optimized design.

\vspace{-.1in}
\subsection{Benchmark Analysis}
In this section we explore performance, the overhead of fault tolerance, while considering the storage capacity from different counter radices all on full application workloads.
\subsubsection{Fault Tolerance Impact on Accuracy}
\label{subsec:fault_tol_acc}

This section shows the impact of CIM faults on applications' accuracy. Count2Multiply, requiring a significantly reduced number of CIM operations, is less susceptible to accuracy degradation. Fig.~\ref{fig:fault} illustrates this in two applications -- DNA filtering and a BERT model -- using a generic MAJ-based RCA implementation, hence serving as a proxy for MAJ-based addition in CIM designs using DRAM or NVMs ~\cite{brackmann2024experimental}. 
As shown, Count2Multiply (\emph{JC}) consistently achieves higher accuracy than RCA-based implementations across all fault probabilities. Moreover, \emph{JC} maintains reliable CIM operation under more severe fault conditions than RCA.

\emph{JC} with protected schemes (\emph{+ECC} or \emph{+TMR}) outperforms RCA due to its lower fault susceptibility from early carry termination.
The MAJ-based \emph{ECC} method (Sec.~\ref{sec:fault-tolerance}) also applies to \emph{RCA} designs and renders more reliable results than using \emph{TMR} for both applications. 
Notably, \emph{TMR}, the fault-tolerance approach used in SOTA~\cite{yuksel2023pulsar}, performs worse than \emph{ECC} with a single repetition.

In DNA filtering, performance degrades more gradually, while BERT exhibits a sharp accuracy drop in task classification on the GLUE dataset ~\cite{wang2018glue} due to its many layers and greater error propagation.
For DNA filtering, an F1 score over 0.9 with a fault rate of 10\% is remarkable and acceptable for less sensitive downstream tasks such as phylogenetic analysis and genome assembly. For the BERT classification task (i.e., MNLI) >70\% accuracy is considered acceptable~\cite{wang2018glue} and is easily achieved for fault rates up to 5\%.

\subsubsection{Fault Tolerance Impact on Performance}
\label{subsec:fault_}
Fig.~\ref{fig:use_cases} compares \emph{C2M} with and without protection, to the baseline SIMDRAM using either workloads.
The impact of the protection scheme is twofold: 1) additional operations are needed to perform \emph{detection} ($7n+7 \rightarrow 13n+16$), and 2) recomputation is required (\emph{correction}) if a fault 
is detected. The detection rate, which informs efficiency overheads, is reported in Tab.~\ref{tab:fault}.
Note that \emph{TMR} would require a \emph{tripling} of operations performed, plus an additional operation to perform majority voting but has the benefit that it eliminates recomputation overheads. Still, as seen in Sec.~\ref{subsec:fault_tol_acc}, TMR exhibits a significantly higher error rate.
For a more intuitive visualization of the overheads from our \emph{ECC} scheme, we present inverted metrics in Fig.~\ref{fig:use_cases}.

Estimating protection overhead requires the memory's fault rate and the number of repetitions to be performed.
In Fig.~\ref{fig:use_cases}, we consider an inherent fault rate of $10^{-4}$ and 1 round of \texttt{FC} computation (repeats = 1). 
This translates to a detected fault rate of $3.5 \times 10 ^{-4}$ per bit (Tab.~\ref{tab:fault}) and $0.16$ per 512-bit row. As shown in Fig.~\ref{fig:use_cases}, the correction overhead in DRAM designs is 19.6\%. This could be improved if a more fine-grained CIM control could repeat only columns with potential errors, decoupling a counter's need to recompute from its distant neighbors.  This has a similar complexity to techniques like differential write for NVMs~\cite{10.1145/1555815.1555759}.

% \vspace{-0.15in}
\subsubsection{Storage Capacity Analysis}

\begin{figure}[tb]
    \centering
    \vspace{-0.15in}
\includegraphics[width=0.75\columnwidth]{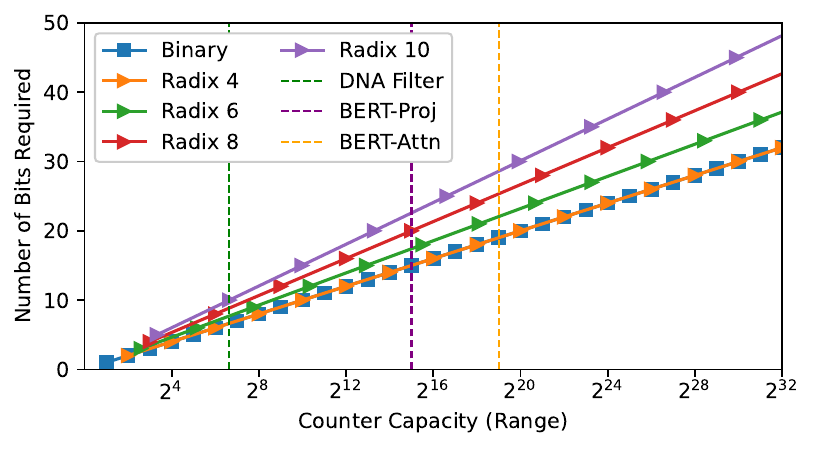}
\vspace{-0.2in}
 \caption{JC capacity and real-world task requirements. }
    \label{fig:storage_cap}
    \vspace{-0.25in}
\end{figure}

While binary encoding (radix 2) is the most efficient in terms of storage capacity, we demonstrated that higher radix counters are better in terms of performance (Fig.~\ref{fig:performance}). 
Here we demonstrate that the storage overhead for high radix counters is moderate in many use cases.
Fig.~\ref{fig:storage_cap} compares the number of bits needed by each counter to achieve the capacity requirements of different applications.
Importantly, the absolute difference in terms of bits does not prohibit high-radix counters.
For example, DNA short-read filtering only requires a capacity of 100 which can be achieved with 10 bits in radix 10 counters or 7 bits in binary. 
BERT projection and attention layers require capacities for accumulating 64 and 792 ternary weight-integer activation products respectively.
Furthermore, our chosen radix-4 counters enjoy the \emph{same density} of storage for a given number of bits as binary encoding.

\section{Conclusions}
\label{sec:conclusion}
We present Count2Multiply, a digital-CIM method that facilitates integer-binary and integer-integer matrix multiplications using high-radix, massively parallel counting with bitwise logic operations. 
Count2Multiply considers reliability a first-class metric and presents a fault tolerance method that is compatible with existing ECC codes, minimizing detection and correction overheads. 
Unlike prior in-DRAM designs, Count2Multiply accelerates sparse matrix operations by skipping zeros, with substantial improvements of the latest DRAM CIM techniques and matching or exceeding GPU performance in many scenarios while providing dramatic energy improvements and better performance per chip area.

\section*{Acknowledgments}
This work was partially funded by the Center for Advancing Electronics Dresden (cfaed) and the German Research Council (DFG) through the HetCIM project (502388442), CO4RTM project(450944241), and the AI competence center ScaDS.AI Dresden/Leipzig in Germany (01IS18026A-D).

\bibliographystyle{IEEEtran}
\bibliography{refs}

\end{document}